\documentclass[sigconf]{acmart}
\usepackage{subcaption}
\usepackage{multirow}
\usepackage{float}

\AtBeginDocument{%
  \providecommand\BibTeX{{%
    \normalfont B\kern-0.5em{\scshape i\kern-0.25em b}\kern-0.8em\TeX}}}

\copyrightyear{2021} 
\acmYear{2021} 
\setcopyright{acmcopyright}\acmConference[AIES '21]{Proceedings of the 2021 AAAI/ACM Conference on AI, Ethics, and Society}{May 19--21, 2021}{Virtual Event, USA}
\acmBooktitle{Proceedings of the 2021 AAAI/ACM Conference on AI, Ethics, and Society (AIES '21), May 19--21, 2021, Virtual Event, USA}
\acmPrice{15.00}
\acmDOI{10.1145/3461702.3462525}
\acmISBN{978-1-4503-8473-5/21/05}

\begin{document}
\fancyhead{} 
\title[Surveilling Surveillance]{Surveilling Surveillance: Estimating the Prevalence of \\Surveillance Cameras with Street View Data}

\author{Hao Sheng}
\email{haosheng@stanford.edu}
\orcid{0000-0002-3386-2825}
\affiliation{%
  \institution{Stanford University}
  \city{Stanford}
  \state{California}
  \country{USA}
  \postcode{94305}
}

\author{Keniel Yao}
\email{keniel.yao@stanford.edu}
\affiliation{%
  \institution{Stanford University}
  \city{Stanford}
  \state{California}
  \country{USA}
  \postcode{94305}
}

\author{Sharad Goel}
\email{scgoel@stanford.edu}
\orcid{0000-0002-6103-9318}
\affiliation{%
  \institution{Stanford University}
  \city{Stanford}
  \state{California}
  \country{USA}
  \postcode{94305}
}

\begin{abstract}
The use of video surveillance in public spaces---both by government agencies and by private citizens---has attracted considerable attention in recent years, particularly in light of rapid advances in face-recognition technology.
But it has been difficult to systematically measure the prevalence and placement of cameras, hampering efforts to assess the implications of surveillance on privacy and public safety.
Here, we combine computer vision, human verification, and statistical analysis to estimate the spatial distribution of surveillance cameras.
Specifically, we build a camera detection model and apply it to 1.6 million street view images sampled from 10 large U.S.\ cities and 6 other major cities around the world, with positive model detections
verified by human experts.
After adjusting for the estimated recall of our model, and accounting for the spatial coverage of our sampled images, we are able to estimate the density of surveillance cameras visible from the road.
Across the 16 cities we consider, the estimated number of surveillance cameras per linear kilometer ranges from 0.2 (in Los Angeles) to 0.9 (in Seoul).
In a detailed analysis of the 10 U.S. cities,
we find that cameras are concentrated in commercial, industrial, and mixed zones,
and in neighborhoods with higher shares of non-white residents---a pattern 
that persists even after adjusting for land use.
These results help inform ongoing discussions on the use of surveillance technology, including its potential disparate impacts on communities of color.
\end{abstract}




\maketitle

\section{Introduction}

Surveillance cameras, also known as closed-circuit television (CCTV) systems, have proliferated in the last several decades as the costs to record and store video have fallen dramatically. 
As of 2016, there were an estimated 350 million surveillance cameras worldwide~\cite{ihs2016}.
The United States, with an estimated 50 million CCTV cameras installed, is believed to have the highest per capita number of surveillance cameras (15.3 CCTV cameras per 100 people) in the world~\cite{precise2019}. 

Past work has found that surveillance cameras may play an important role in crime prevention and investigation, but there is also growing concern about the dangers cameras pose to privacy and equity.
Further, recent advances in facial recognition technology significantly amplify both the potential costs and the potential benefits of widespread surveillance,
as it is now possible to identify and track specific individuals across space and time.
While these technical advances promise to aid law enforcement efforts, they may also unjustly concentrate policing on more heavily monitored communities.
This surveillance may also hinder longstanding freedoms of speech and association, as it becomes easier to identify those participating in public gatherings, potentially dissuading dissent.

Despite the wide-ranging implications of surveillance cameras on public safety, police enforcement, and democratic governance, relatively little is known about the precise number and placement of cameras, hampering efforts to assess their impacts.
Past work to gauge the prevalence and spatial distribution of surveillance cameras has either examined aggregate production or shipping numbers, or relied on public disclosures in select jurisdictions---approaches that suffer from limitations of scale and scope.

To address these limitations, 
\citet{turtiainen2020cctv} note that researchers could, in theory, map surveillance cameras by applying computer vision algorithms to street view data, which provide nearly complete visual coverage of many cities.
Building on that insight, here we describe and implement a scalable method for measuring the distribution of outdoor surveillance cameras across the United States, and, more generally, across the world.
Specifically, we couple computer vision algorithms with a verification pipeline by expert human annotators, together with statistical adjustment, to analyze a large-scale corpus of street view images.
In this manner, we leverage the proliferation of cameras and image data themselves to quantify the prevalence of surveillance technology.

To carry out this analysis, we use the public repository of images collected as part of Google's Street View service, launched in 2007. Since its inception, millions of 360-degree panoramas have been collected by cameras mounted on the roof rack of Google Street View cars, covering more than 10 million miles across 83 countries~\cite{raman2017}. 
The rich archive of historical street view images provides opportunities to understand the evolution of the built environment, particularly the adoption of surveillance cameras on a global scale.  
However, it is still extremely challenging---if not impossible---for humans to eyeball millions of images and spot cameras from the diverse street view context: a camera usually consists of 30--50 pixels out of over 400,000 pixels in a standard 640 $\times$ 640 street view image. 

To scour this collection of images, we train and apply a computer vision algorithm to first filter street view data to those candidate images likely to contain a surveillance camera.
We specifically start with a random selection of 1.6 million street view images from 10 large U.S. cities and 6 other major cities, which contains approximately 6,000 positive model detections.
This curated set of candidate images is then examined by
human experts for verification. 
To go from verified camera detections in our sample to city-wide estimates, we further
estimate both the recall of our model (which we find is 0.63), and 
the proportion of the city covered by our sample.
This latter quantity is computed 
based on the recorded camera position and angle, coupled with high-precision data on the road network and building footprints.

We find substantial variation in the density of visible surveillance cameras across the 16 cities we consider, ranging from 0.07 cameras per linear kilometer along the road network in Seattle to 0.95 cameras per kilometer in Seoul.
Examining the 10 U.S. cities in greater detail, we find that surveillance cameras are concentrated in 
commercial, industrial, and mixed city zones,
and also in areas with higher shares of non-white residents.
This concentration of cameras in majority-minority neighborhoods persists even after adjusting for zone, pointing to the potential disparate impacts of surveillance technology on communities of color.

\section{Related work}
Our work connects to several interrelated strands of research in computer vision, urban computing, and privacy, which we briefly summarize below.

\subsection{Street View Understanding}
Visual scene understanding~\cite{hoiem2015guest} is one of the most fundamental and challenging goals in computer vision.
In part because of its potential to support self-driving vehicles, both the industrial and scientific communities have put considerable effort and investment into designing and creating labeled street view datasets for training and evaluating deep learning models, such as CamVid~\cite{brostow2008segmentation}, the KITTI Vision Benchmark Suite~\cite{geiger2013vision}, Cityscapes~\cite{cordts2016cityscapes}, and Mapillary Vistas~\cite{neuhold2017mapillary}. 
Based on these datasets, several studies exploit the characteristics of urban-scene images and propose object segmentation~\cite{choi2020cars,liu2015layered} and change detection~\cite{alcantarilla2018street} algorithms for general street view understanding.

Related research has focused on detecting specific elements from street images, including greenery~\cite{li2015assessing}, buildings~\cite{kang2018building}, and city infrastructure such as utility poles~\cite{zhang2018using}.
Of particular relevance to our work, 
\citet{neuhold2017mapillary} built an image segmentation model to identify---among other objects---CCTVs in street view data.
The publicly available \citeauthor{neuhold2017mapillary} Mapillary Vistas Dataset contains over 20,000 labeled images but fewer than 100 labeled cameras, leading to relatively poor performance on the specific task of detecting cameras.
More recently,
\citet{turtiainen2020cctv} developed a state-of-the-art object detection model tailored specifically to CCTVs,
based on nearly 10,000 images of cameras that they collected and labeled.
That dataset, however, has not been publicly released at the time of writing.
As a result, we constructed (and have released) our own labeled camera dataset and built a camera detection model using standard computer vision techniques. 

\subsection{Urban Computing}
Urban computing aims to tackle major issues in cities---such as traffic control, public health, and economic development---by modeling and analyzing urban data. 
A large body of research has shown that it is possible to infer socioeconomic information from satellite images~\cite{jean2016combining, Sheng_2020_CVPR_Workshops}, monitor human mobility~\cite{xu2018human}, and identify geo-tagged social network activities~\cite{schwartz2014social}. 
Recent studies using street view images have dramatically increased the accuracy of processed data, as well as the geographic resolution analyzed.  
By manually scoring street view images from 2,709 city blocks, \citet{hwang2014divergent} find that gentrification in Chicago from 2007 to 2009 was negatively associated with the concentration of minority groups. 
\citet{mooney2016use} labeled the characteristics of 532 intersections in New York City, such as curb cuts and crosswalks, to assess environmental contributions to pedestrian injury. 

As an alternative to relying on human experts to annotate street view images, modern computer vision algorithms have a much higher throughput with close to zero cost, enabling researchers to scale to multiple cities. 
For example, \citet{gebru2017using} enumerated 22 million automobiles (8\% of all vehicles in the U.S.) in 50 million street view images to accurately
estimate local income, race, education, and voting patterns. 
In our work, we draw on the merits of both approaches, combining high-recall computer vision algorithms with high-precision human verification in a unified estimation pipeline. 
\begin{figure*}[ht!]
  \centering
  \includegraphics[width=5in]{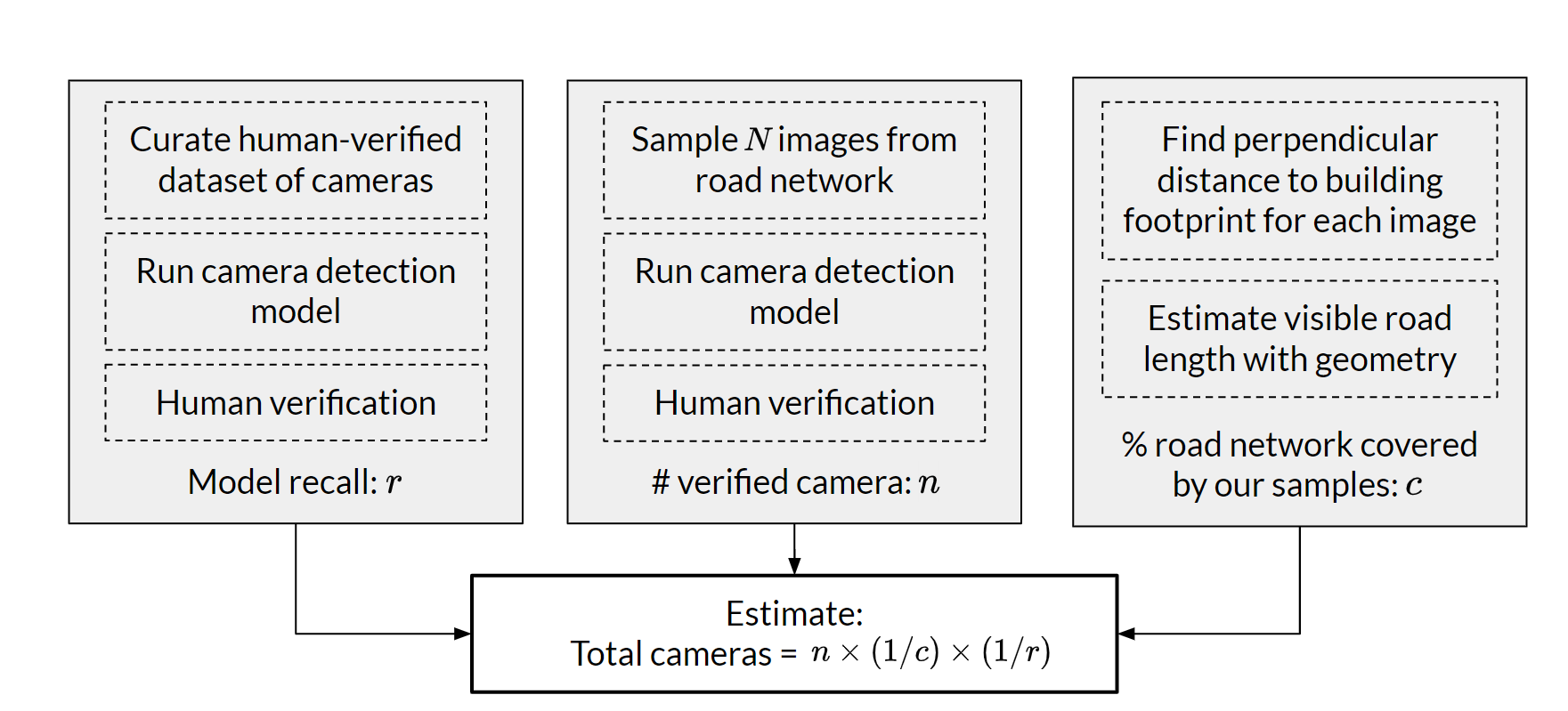} 
  \caption{\textbf{Camera estimation workflow.} We take a three-step approach to estimating the prevalence and spatial distribution of surveillance cameras across 16 major cities. 
  First, we create a labelled dataset of street view images to estimate the recall of our camera-detection algorithm.
  Next, we run this camera-detection algorithm on a random sample of street images,
  and then use human annotators to identify images with actual cameras.
  Finally, we estimate the proportion of the road network covered by our random sample of street images.
  }
  \label{fig:flowchart}
\end{figure*}
\subsection{Surveillance and Privacy}
 While past work has found that surveillance cameras play an important role in crime investigation~\cite{king2008citris} and deterrence~\cite{welsh2015effectiveness}, cameras also pose significant challenges to privacy. 
Legal scholars have long considered the ramifications of cameras on First Amendment freedoms and the constitutional right to privacy~\cite{robb1980police}. 
More recently, scholars have been concerned with the role of surveillance cameras in predictive policing \cite{joh2016discretion}, 
in enabling the adverse effects of facial recognition and computer vision~\cite{stanley2019robot,calo2010fake,buolamwini2018shades,nkonde2020black}, 
and with the threat of surveillance hacking~\cite{hermann2018hack,quintin2015license}. 
These concerns have led to bans on facial-recognition technology by law enforcement in San Francisco, Boston, and Portland~\cite{banfacialrecognition},
as well as the drafting of federal legislation~\cite{frtbill}.

Despite these concerns, there has been limited success in identifying the number and geospatial distribution of cameras. 
The Electronic Frontier Foundation (EFF) recently acquired the locations of cameras accessible by prosecutors in San Francisco~\cite{maass2019camera}. 
Other private-market researchers have estimated the prevalence of installed cameras at a national level through unit shipments~\cite{jenkins2019surveil}. 
However, neither of these approaches are able to estimate the prevalence and specific locations of public and private cameras at scale, hindering downstream analysis on the impacts of surveillance.

\section{Data and Methods}

For 16 major cities, we estimate the total number and spatial distribution 
of surveillance cameras visible from the street.
We specifically consider the 10 cities with the highest urban density in the U.S., among those with at least 500,000 residents, and 6 other major cities in Asia and Europe. 
Our statistical estimation procedure involves three key steps.
First, we compile a dataset of street view images both with and without cameras and label these images with segmentation masks. 
We then train a camera segmentation model on this dataset, 
and, importantly, estimate the 
recall of our detection algorithm on a held-out validation dataset.
Second, we run our camera detection algorithm on a random sample of street view images.
All positive camera detections are then reviewed by human experts to remove false positives. 
Finally, by combining the geometry of the camera angle, 
the road network, and building footprints, 
we calculate our sample's coverage of the road network.
These three steps are outlined in
Figure~\ref{fig:flowchart}.
In the following sections, we describe the data used in our analysis and more fully detail each step in our estimation pipeline.

\subsection{Data}
\begin{table}[!htbp] 
  \centering
    \resizebox{1\columnwidth}{!}{%
        \begin{tabular}{@{\extracolsep{5pt}} lrrr} 
        \\[-1.8ex]\hline 
        \toprule
        City & Population & Area (sq. km) & Road length (km) \\ 
        \midrule
        Los Angeles & 3,793,000 & 1,213 & 21,095 \\ 
        New York City & 8,175,000 & 783 & 16,362 \\ 
        Chicago & 2,696,000 & 589 & 10,449 \\ 
        Philadelphia & 1,526,000 & 347 & 6,759 \\ 
        Seattle & 609,000 & 217 & 5,569 \\ 
        Milwaukee & 595,000 & 248 & 4,899 \\ 
        Baltimore & 621,000 & 209 & 3,746 \\ 
        Washington, D.C. & 602,000 & 158 & 3,262 \\ 
        San Francisco & 805,000 & 121 & 3,101 \\ 
        Boston & 618,000 & 125 & 2,589 \\ 
        \midrule
        Tokyo & 13,159,000 & 2,194 & 46,688 \\ 
        Bangkok & 8,305,000 & 1,569 & 34,692 \\ 
        London & 8,174,000 & 1,572 & 28,907 \\ 
        Seoul & 9,630,000 & 605 & 14,748 \\ 
        Singapore & 3,772,000 & 728 & 5,794 \\ 
        Paris & 2,244,000 & 106 & 1,853 \\ 
        \bottomrule
        \end{tabular} 
        }
    \caption{We estimate the prevalence and location of surveillance cameras in the 10 large U.S.\ cities and 6 other major cities around the world. Above we list the population and land area of these cities, and the estimated length of the road network within the city bounds, in descending order of road length.}
    \label{tab:citystats}
\end{table}

\begin{figure}[ht!]
  \centering
  \includegraphics[width=0.2\textwidth]{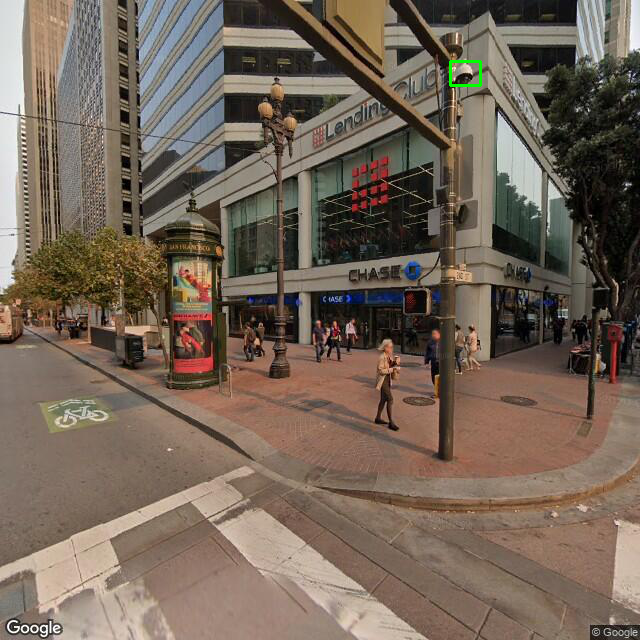} 
  \includegraphics[width=0.2\textwidth]{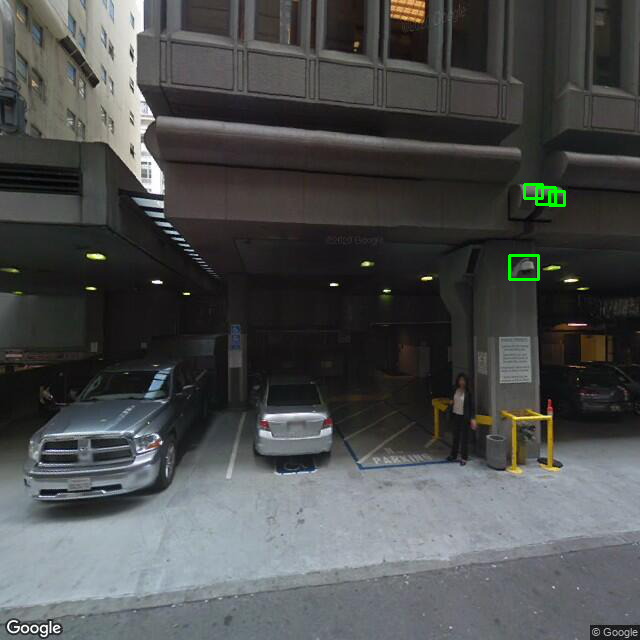} 
  \includegraphics[width=0.2\textwidth]{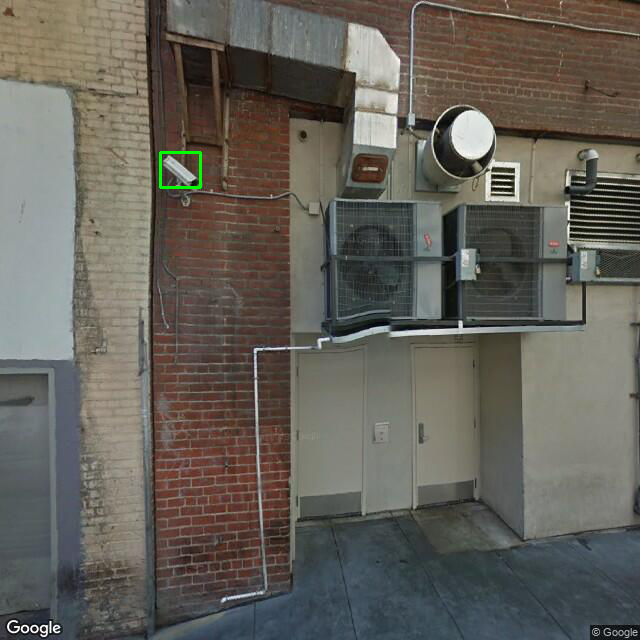} 
  \includegraphics[width=0.2\textwidth]{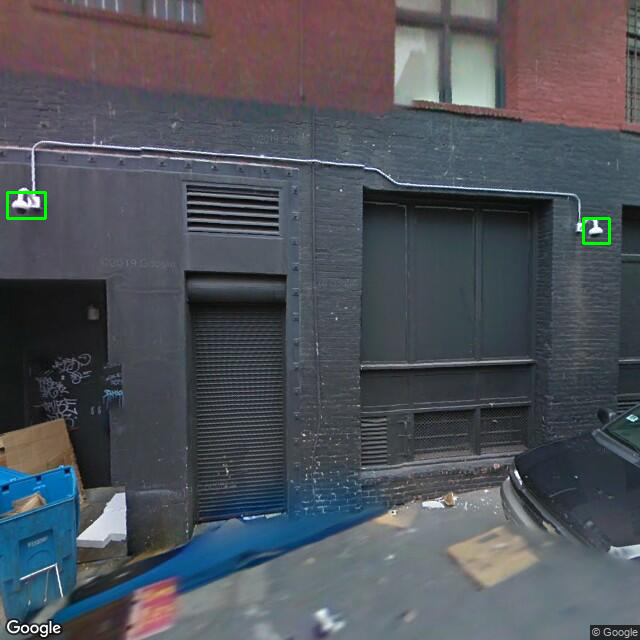} 
  
  \caption{Examples of surveillance cameras in San Francisco. \textbf{Upper left}: Dome camera mounted on a traffic pole. \textbf{Upper right}: Dome camera mounted on the wall of a parking structure. \textbf{Lower left}: A wall mounted directed camera. \textbf{Lower right}: Two wall-mounted dome cameras. }
  \label{fig:example}
\end{figure}
We analyze the 16 cities listed in Table~\ref{tab:citystats}.
For each city, we obtained the road network and building footprints from OpenStreetMap~\cite{OpenStreetMap,boeing2017osmnx}. 
U.S.\ Census maps were used to restrict the geospatial data to the city's administrative borders.
All street view images used for model training and camera detection were accessed through the Google Static Streetview API.\footnote{\url{https://developers.google.com/maps/documentation/streetview}}
We further used San Francisco camera location data from the EFF~\cite{maass2019camera} to construct training and evaluation datasets for our model.

\subsection{Step 1: Model Training and Evaluation}

We start by creating training and evaluation datasets for our camera detection model. 
For each of the 2,660 geo-tagged cameras in San Francisco identified by the EFF, 
we pulled the closest street view images from 2012--2019 (if there is a scene available within 30 meters). 
Manually labeling the resulting 13,240 images yielded 861 positive instances containing 977 cameras.
We note that many of the cameras listed in the EFF dataset appear to be indoors or otherwise are not visible from the street. 
In Figure~\ref{fig:example}, we show several labeled examples. 

We frame our camera detection problem as a binary image segmentation problem
to maximize learning from a limited number of samples.
We split the positive images by location into 70\%/15\%/15\% training/validation/test sets, making sure images from the same site always belong to the same split. We further include all camera instances from Mapillary Vista into our training dataset. By mixing with the negative images, we end up with 5,298 images for training, 1,040 for validation, and 1,040 for testing. 

\begin{figure}[ht!]
  \centering
  \includegraphics[width=0.45\textwidth]{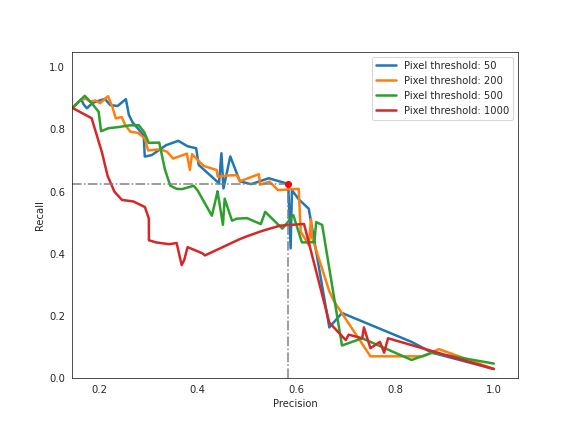} 
  \caption{Model precision and recall at camera instance level as we vary the classification threshold, for different size filters (in pixels). We decided to use a probability threshold of 0.75 and a size threshold of 50 pixels, which yields precision and recall equal to 0.58 and 0.63, respectively.}
  \label{fig:performance}
\end{figure}

For ease, we use off-the-shelf methods to train our computer vision model (for state-of-the-art camera detection, see \citet{turtiainen2020cctv}). 
In particular, our segmentation model follows the architecture of DeepLab V3+~\cite{chen2017rethinking,chen2018encoder} with an EfficientNet-b3~\cite{tan2019efficientnet} backbone. 
We apply a random horizontal flip and randomly crop the original image (640 x 640 pixels) to 320 x 320 before feeding it into the model during training.
In the inference phase, we first crop the input image into four patches and then merge the output segmentation maps back to the original size. 
The segmentation model's performance is shown in Table~\ref{tab:performance}.
To aggregate the pixel level prediction to the instance level, we first apply a morphological dilation with a 3x3 kernel to merge detected areas and filter false detections by size.  After validating several combinations of pixel-level probability thresholds and size filters,
we decided to use a probability threshold of 0.75 and a size threshold of 50 pixels, which yields precision and recall equal to 0.58 and 0.63, respectively (see Figure~\ref{fig:performance}).

\begin{figure}[t]

  \centering
  \includegraphics[width=0.2\textwidth]{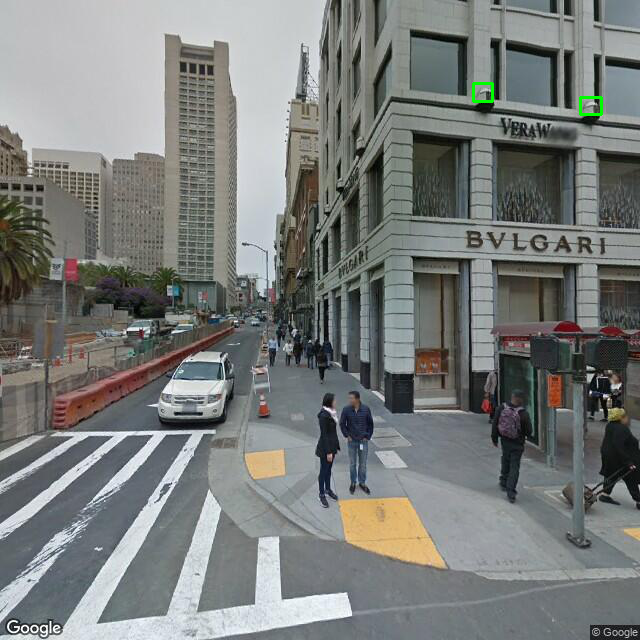} 
  \includegraphics[width=0.2\textwidth]{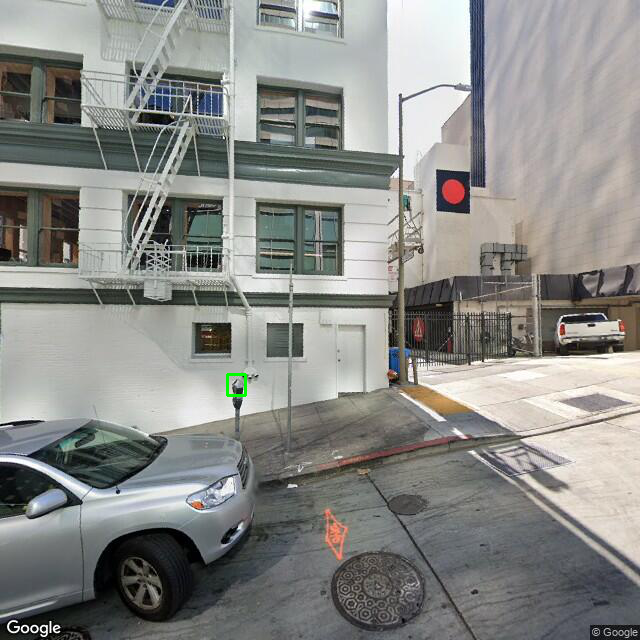} 
  \includegraphics[width=0.2\textwidth]{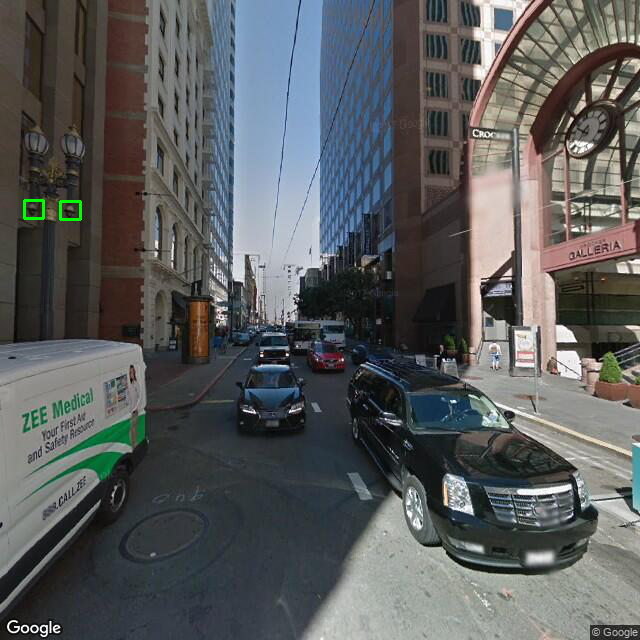}
  \includegraphics[width=0.2\textwidth]{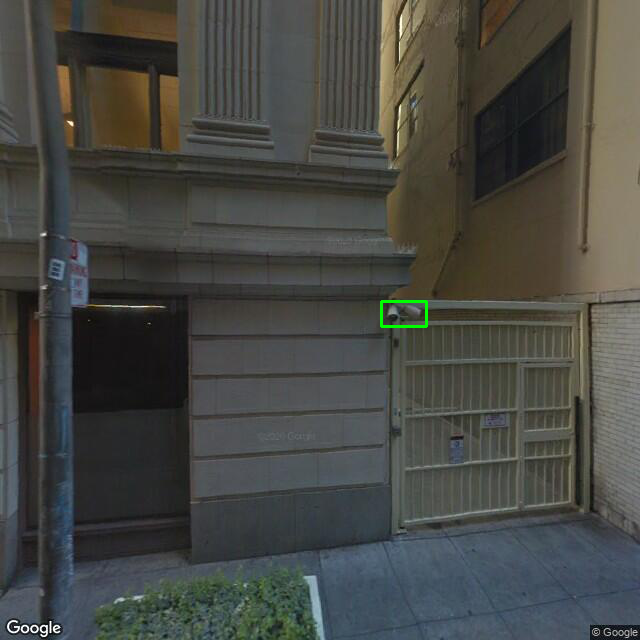}

  \caption{Examples of false positives and other errors from our camera detection model. 
  \textbf{Upper left}: Building structure.
  \textbf{Upper right}: Parking meter.
  \textbf{Lower left}: Street lamp.
  \textbf{Lower right}: Our algorithm incorrectly merged two cameras into one detection. 
  }
  \label{fig:failure}
\end{figure}

In Figure~\ref{fig:failure}, we present several illustrative failures of our detection model. The model is occasionally confused by objects that share some of the visual features of cameras, such as building structures, parking meters, and street lamps. In some instances, our model also merges multiple cameras into one detection. These problems are mitigated by the human verification step, as described in the next section. 
\begin{table}[t]
    \centering
    \resizebox{0.8\columnwidth}{!}{
        \begin{tabular}{l|ccc}
            \toprule
            Data Split  & IoU$_{\text{camera}}$ & Accuracy & F1-score \\
            \midrule
            Validation Set & 0.71 & 0.94  & 0.90\\
            Test Set       & 0.69 & 0.93  & 0.89\\
            \bottomrule
        \end{tabular}
    }
    \caption{Performance of our camera segmentation model in pixel space.}
    \label{tab:performance}
\end{table}

\subsection{Step 2: Camera Detection and Verification}
\begin{figure*}[ht!]
  \centering
  \includegraphics[width=0.3\textwidth]{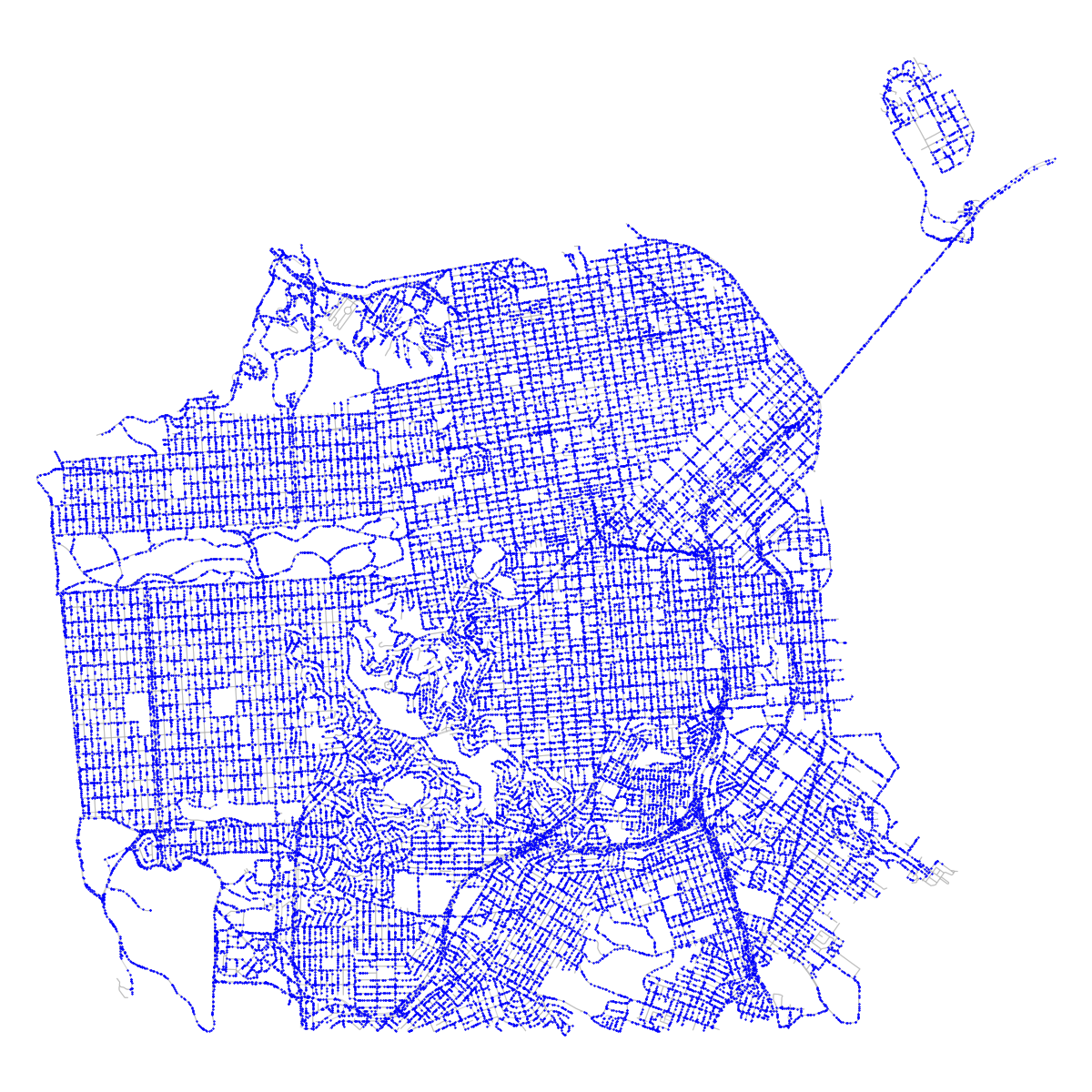}
  \includegraphics[width=0.3\textwidth]{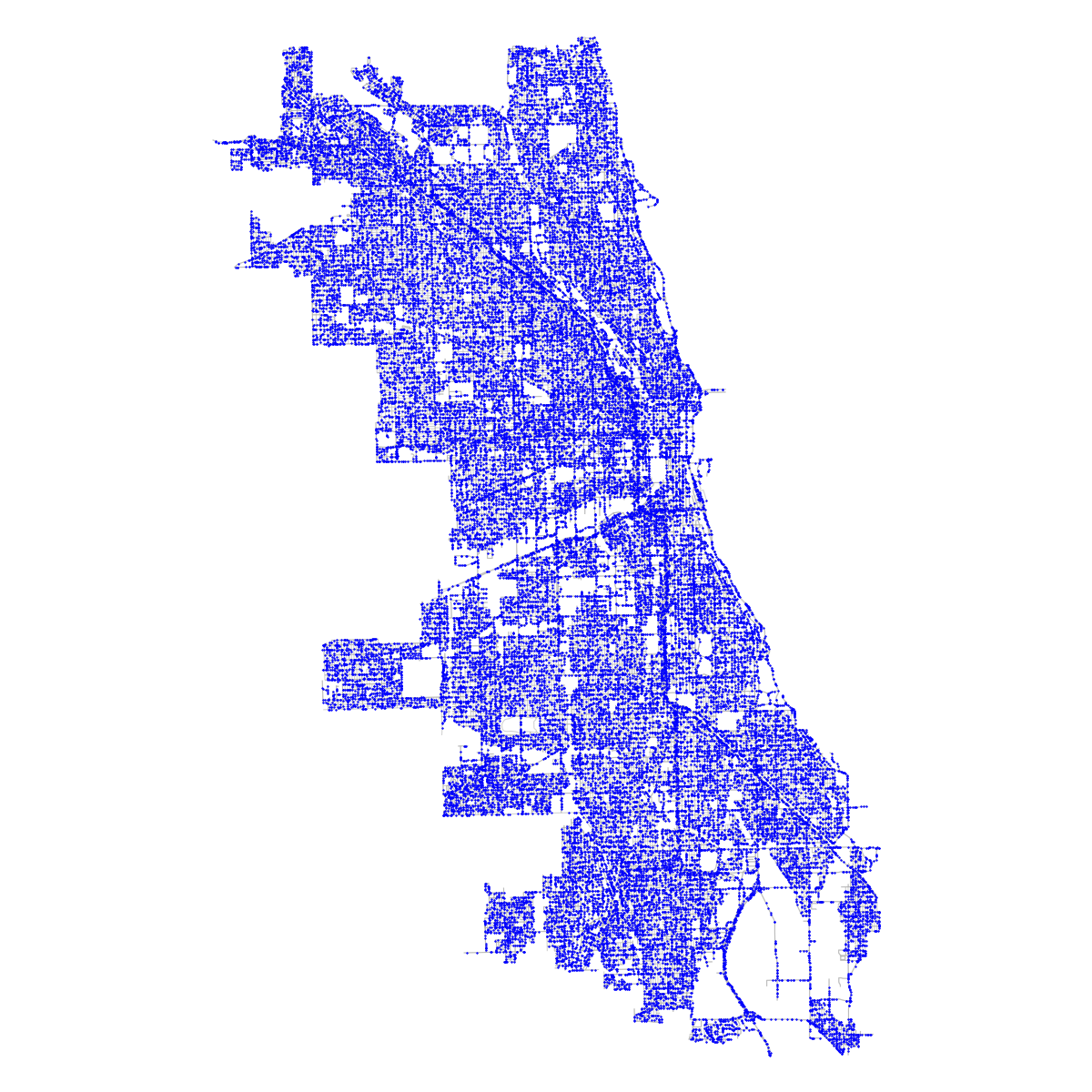}
  \includegraphics[width=0.3\textwidth]{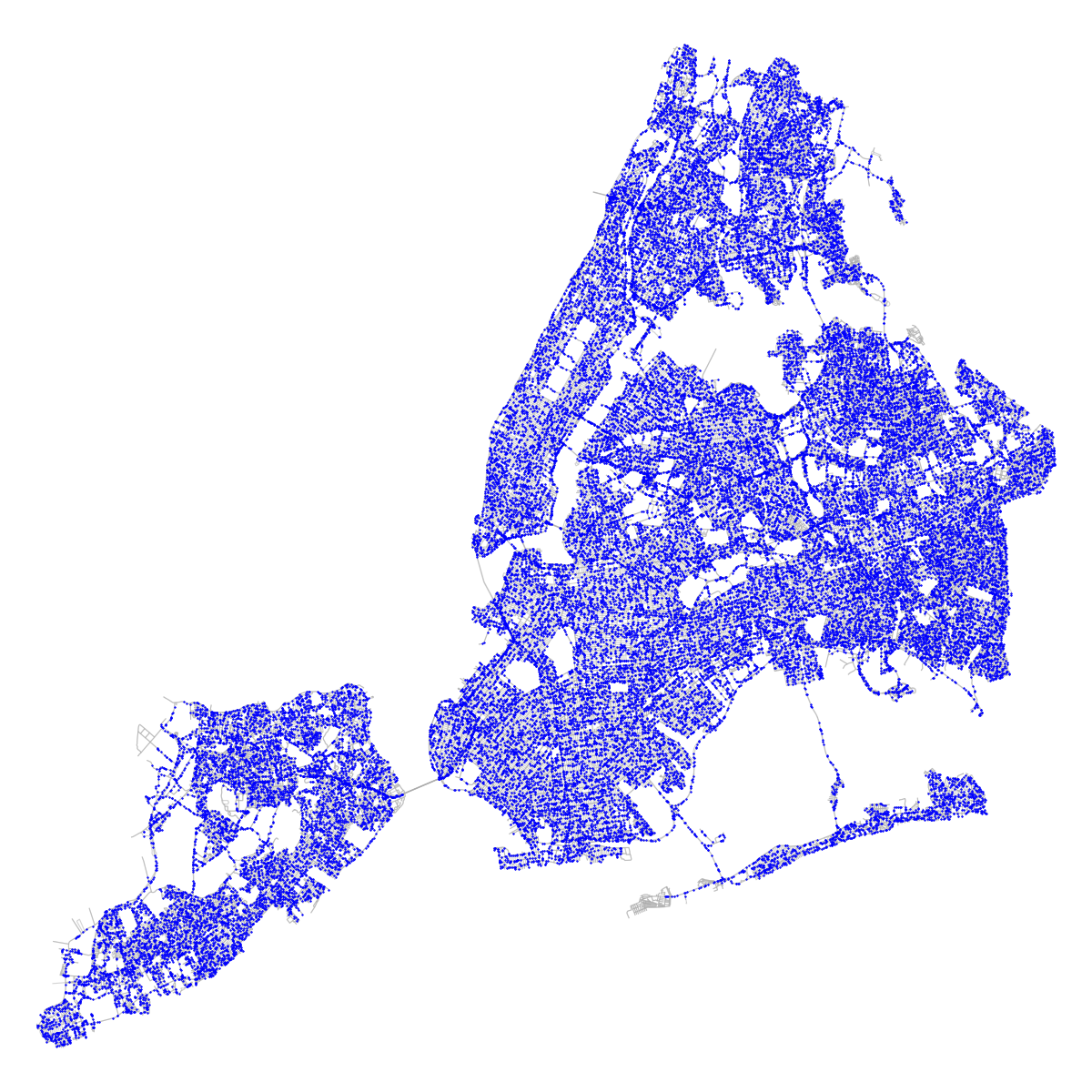} 
  \caption{The spatial distribution of sampled points on the road network for three illustrative cities. \textbf{Left}: San Francisco; \textbf{Middle}: Chicago; \textbf{Right}: New York City.}
  \label{fig:spatial}
\end{figure*}
\begin{figure}[ht!]
  \centering
  \includegraphics[width=0.47\textwidth]{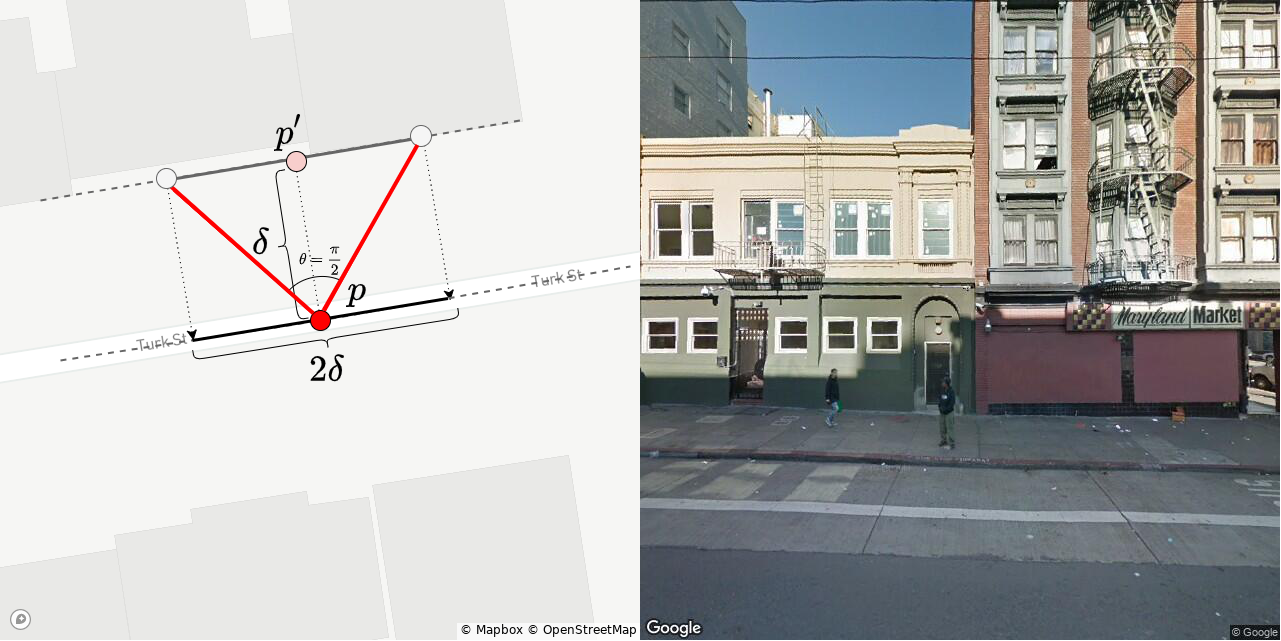} 
  \caption{Illustration of how we estimate the road segment coverage for one image.
  \textbf{Right}: The Google Street View image. 
  \textbf{Left}: Corresponding field of view on map. 
  We obtain the point location of each image, $p$, and find the closest point $p'$ within the footprint of the nearby buildings. The length of the road in view is twice the distance $d$ between $p$ and $p'$, as each image has a 90-degree field facing the road. We estimate the road segment covered in the shown image has a length of $19.55$ meters.}
  \label{fig:geometry}
\end{figure}
\begin{figure*}[ht!]
  \centering
    \begin{subfigure}[t]{.22\linewidth}
    \centering 
    \includegraphics[width=1\textwidth]{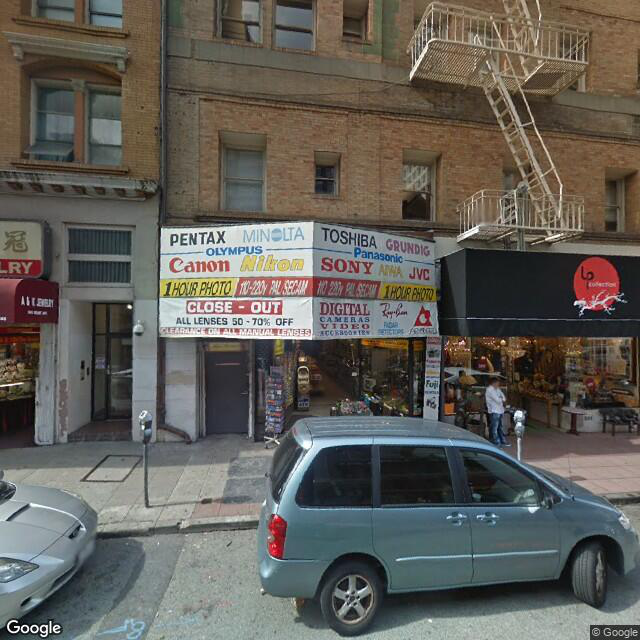} 
    \caption{Raw Image}
   \end{subfigure}
   \begin{subfigure}[t]{.22\linewidth}
    \centering 
    \includegraphics[width=1\textwidth]{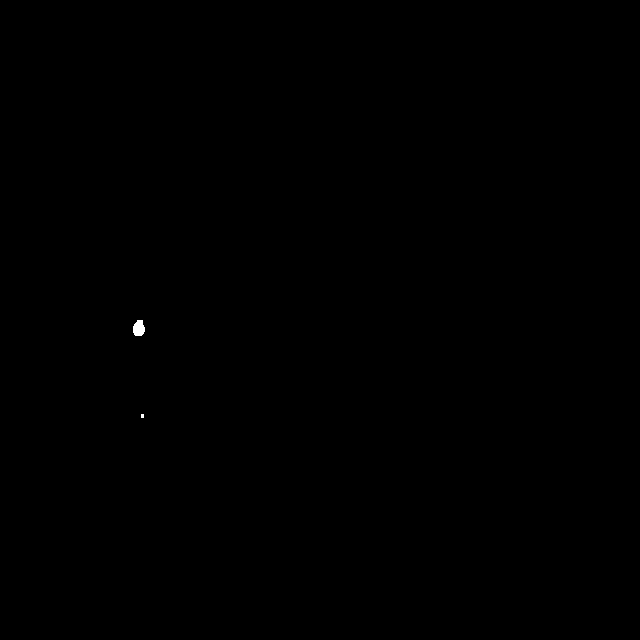} 
    \caption{Segmentation}
   \end{subfigure}
   \begin{subfigure}[t]{.22\linewidth}
    \centering 
    \includegraphics[width=1\textwidth]{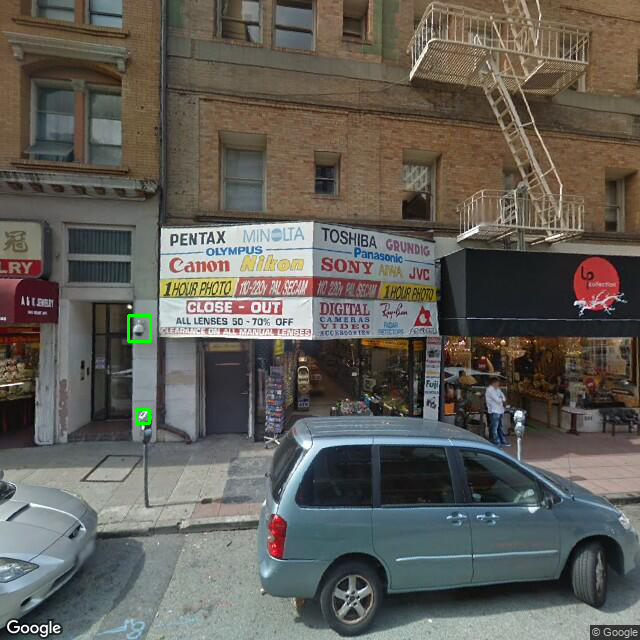} 
    \caption{Bounding Boxes}
   \end{subfigure}
   \begin{subfigure}[t]{.22\linewidth}
    \centering 
    \includegraphics[width=1\textwidth]{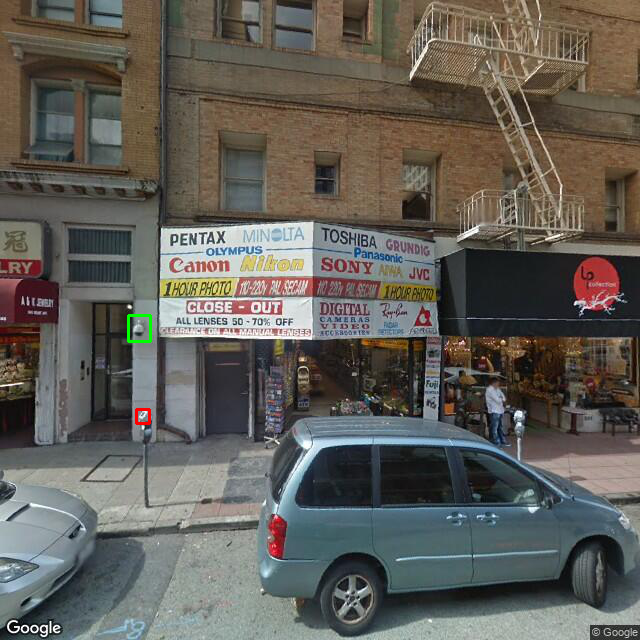} 
    \caption{Verification}
   \end{subfigure}
  \caption{Illustration of our machine detection and human verification pipeline. 
  For each image yielding a positive model detection, human coders verify the detection based on the raw image (a) and bounding boxes (c) generated from the segmentation (b). In this example, the human coder confirmed the upper box is a camera but not the lower box (d).}
  \label{fig:verification}
\end{figure*}
For each city, we sampled street view images at $N=100,000$ points chosen uniformly at random from the road network.\footnote{For reference, there are more than 400,000 points covered with distinct street view panoramas in San Francisco.}
For approximately 3\% of the selected points, there was
no street view coverage within 10 meters, in which case we discarded and then re-sampled the location.
Figure~\ref{fig:spatial} shows the spatial distribution of the sampled points for three example cities: San Francisco, New York, and Chicago. 

For each location, we then selected a 360-degree street view panorama.
For London, Paris, and the 10 American cities, we selected the oldest available image taken between 2015 and 2021;
for the remaining cities, we selected the oldest available image in the Google Maps corpus, which goes back to 2007.
We note that this sampling strategy is the result of a coding error; our intention was to select the \emph{newest} available image at each location.
Finally, for each location sampled, we randomly selected one out of the two 90-degree views with a midpoint perpendicular to the orientation of the road (see Figure~\ref{fig:geometry}). This approach provides the maximum view of the roadside.

We ran our camera detection model on the resulting set of 100,000 images for each of the 16 cities.
Annotators received the raw image and bounding boxes highlighting the predicted cameras, which were automatically generated from the model segmentation outputs. This process yielded 6,281 positive images with a total of 6,469 camera detections, 
all of which were then verified by a human annotator. 

Figure~\ref{fig:verification} illustrates the pipeline from the raw image to segmentation and bounding boxes to human verification.
In our subsequent analysis, we only consider these human-verified camera detections.

\begin{figure*}[htbp!]
    \makeatletter
    \DeclareRobustCommand{\onontimes}{%
      \mathbin{\mathpalette\on@ntimes\relax}%
    }
    \newcommand{\on@ntimes}[2]{
      \vcenter{\hbox{
        \sbox0{\m@th$#1\otimes$}
        \setlength\unitlength{\wd0}
        \begin{picture}(1,1)
        \linethickness{0.35pt}
        \put(.5,.5){\circle{.8}}
        \end{picture}
      }}
    }
    \makeatother
  
  \begin{subfigure}[t]{.19\linewidth}
      \centering
      \includegraphics[width=1\textwidth]{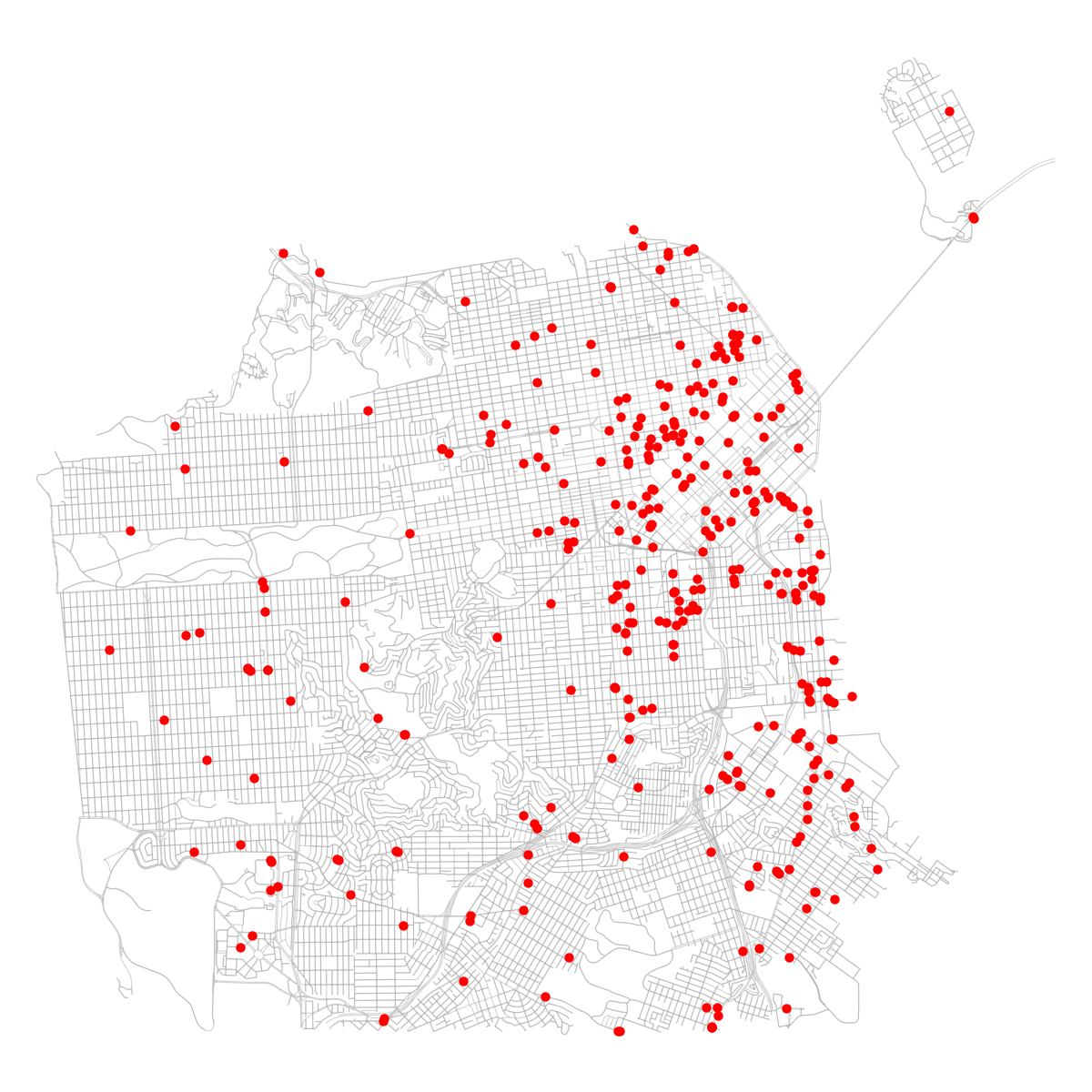} 
      \caption{San Francisco}
  \end{subfigure}
  \begin{subfigure}[t]{.19\linewidth}
      \centering
      \includegraphics[width=1\textwidth]{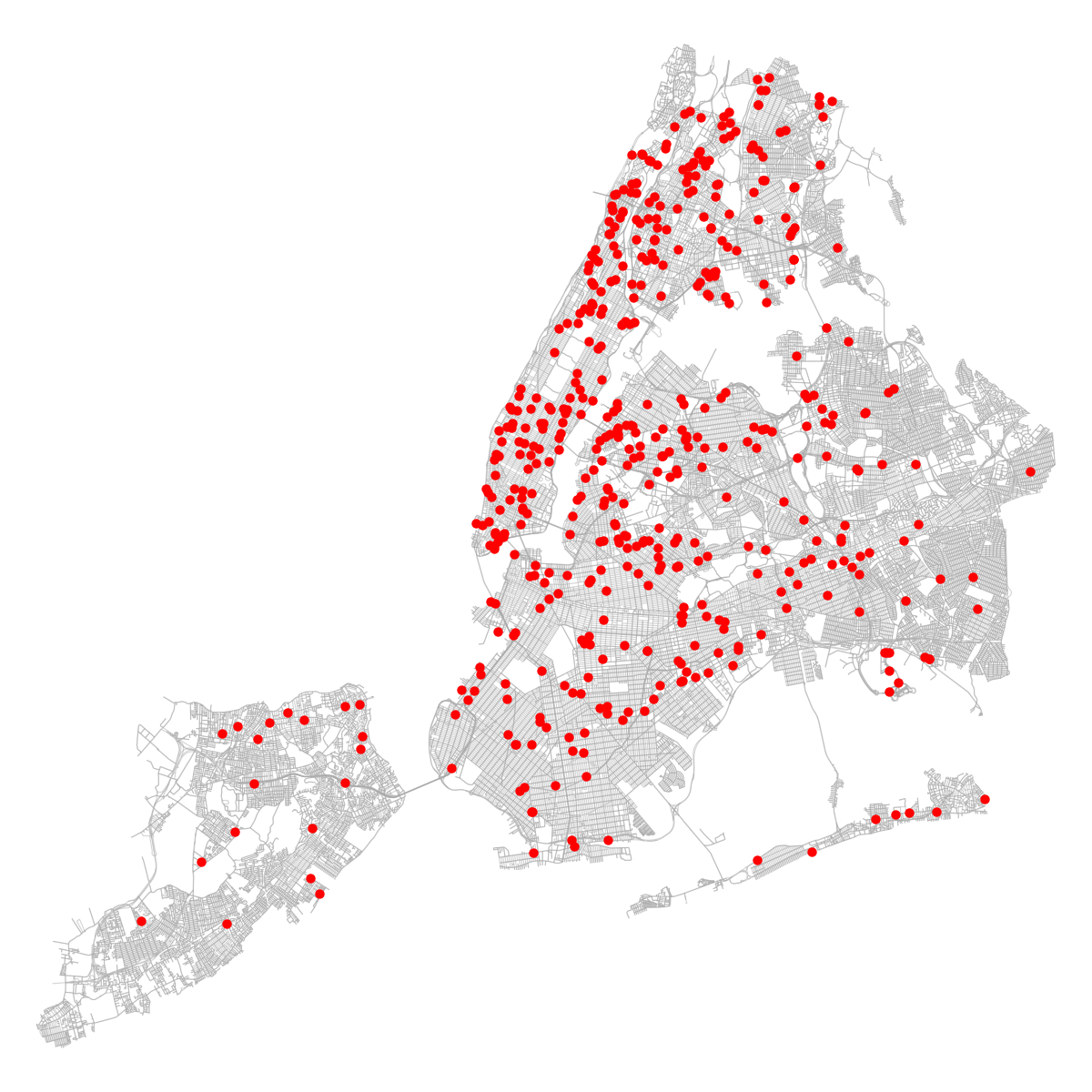} 
      \caption{New York City}
  \end{subfigure}
  \begin{subfigure}[t]{.19\linewidth}
    \centering 
    \includegraphics[width=1\textwidth]{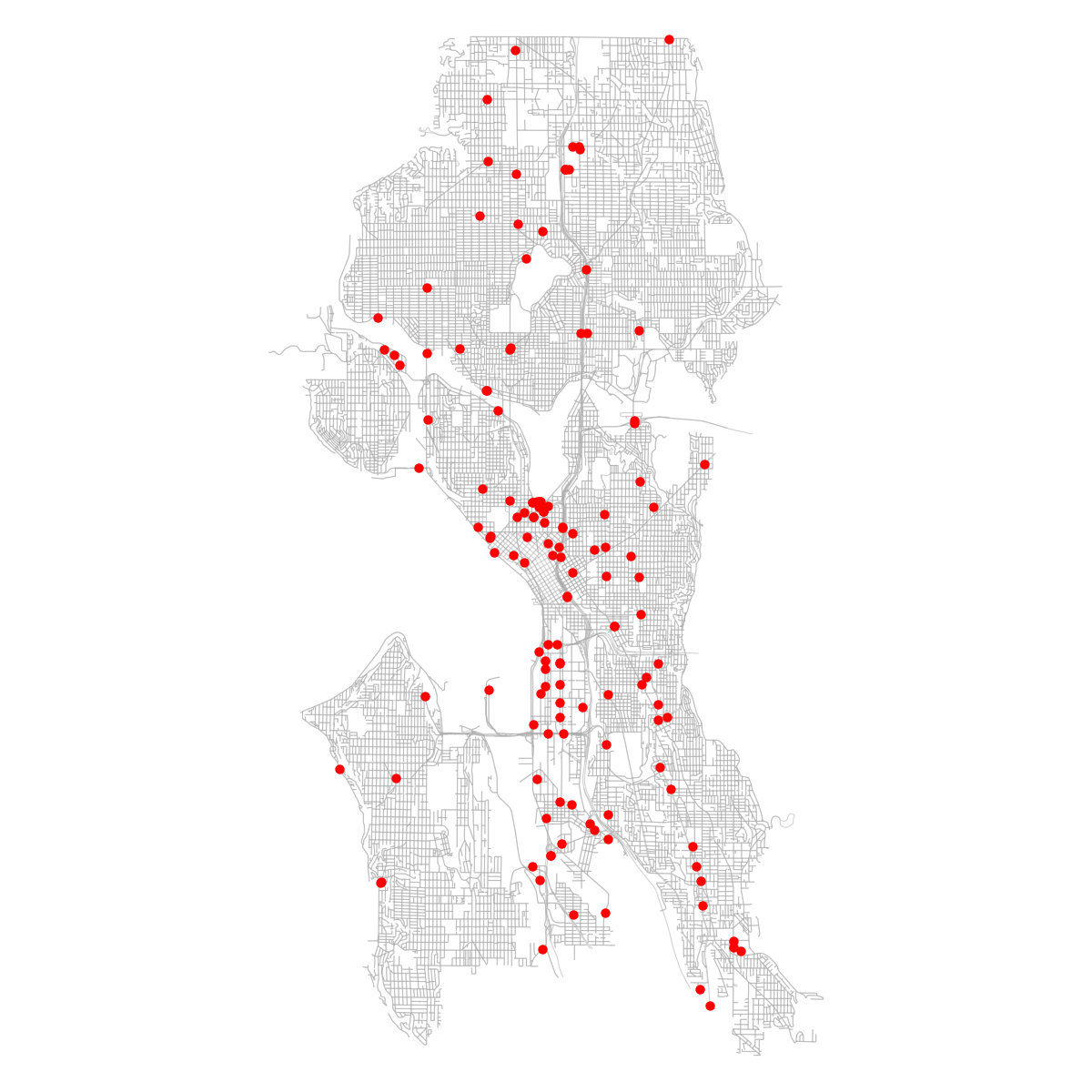}
    \caption{Seattle}
  \end{subfigure}
    \begin{subfigure}[t]{.19\linewidth}
    \centering 
    \includegraphics[width=1\textwidth]{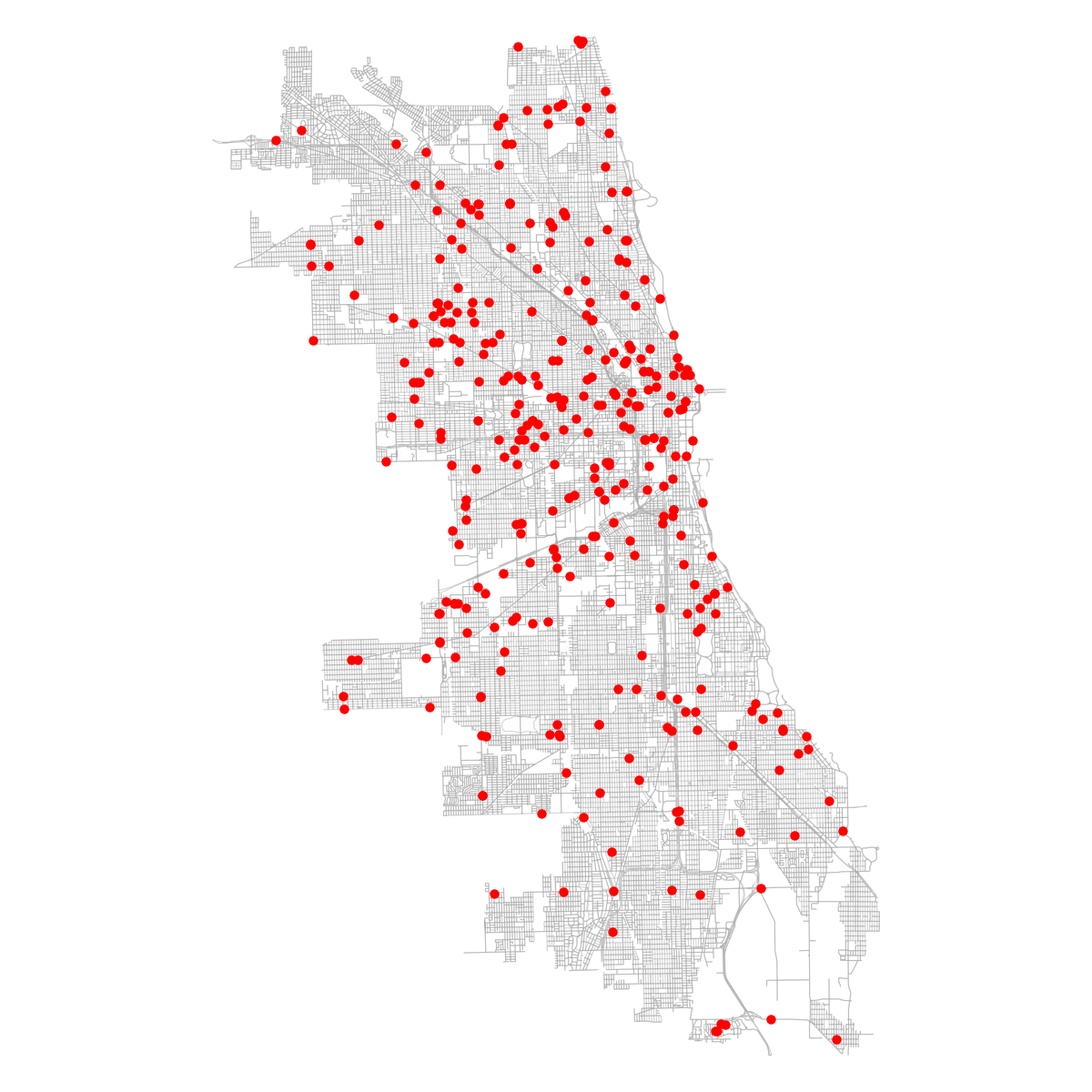}
    \caption{Chicago}
  \end{subfigure}
    \begin{subfigure}[t]{.19\linewidth}
    \centering 
    \includegraphics[width=1\textwidth]{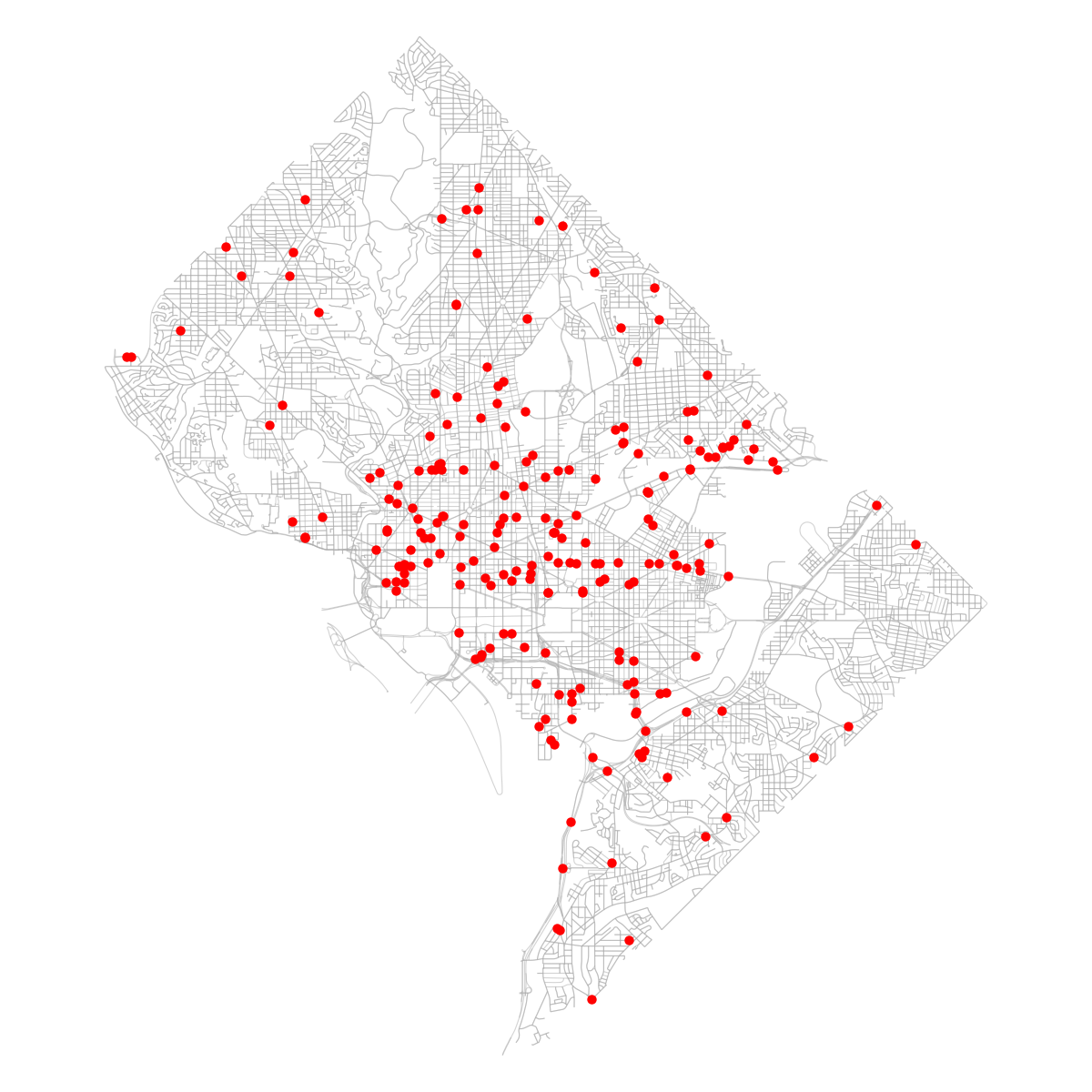} 
    \caption{Washington, D.C.}
  \end{subfigure}
    \begin{subfigure}[t]{.19\linewidth}
    \centering 
    \includegraphics[width=1\textwidth]{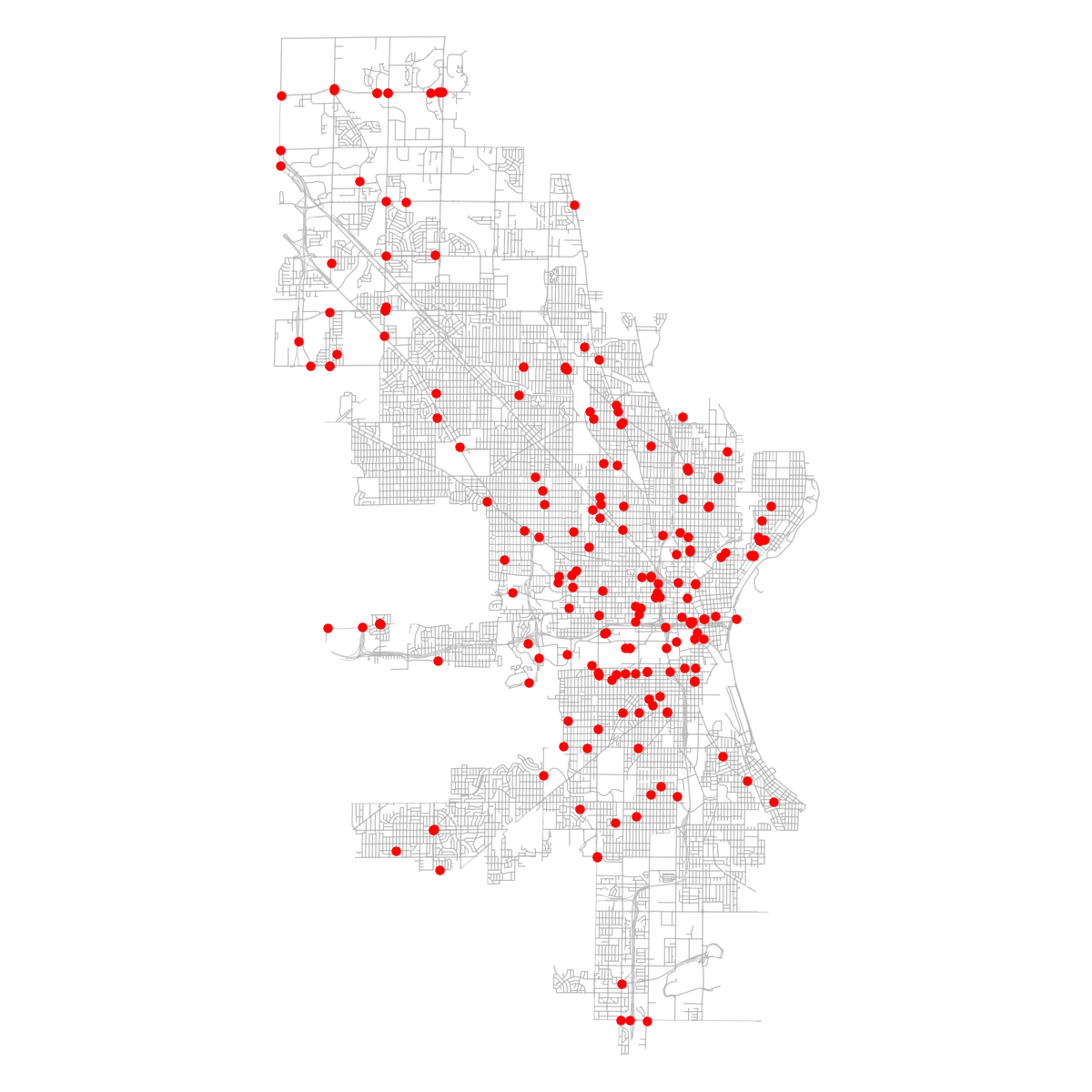} 
    \caption{Milwaukee}
  \end{subfigure}
    \begin{subfigure}[t]{.19\linewidth}
    \centering 
    \includegraphics[width=1\textwidth]{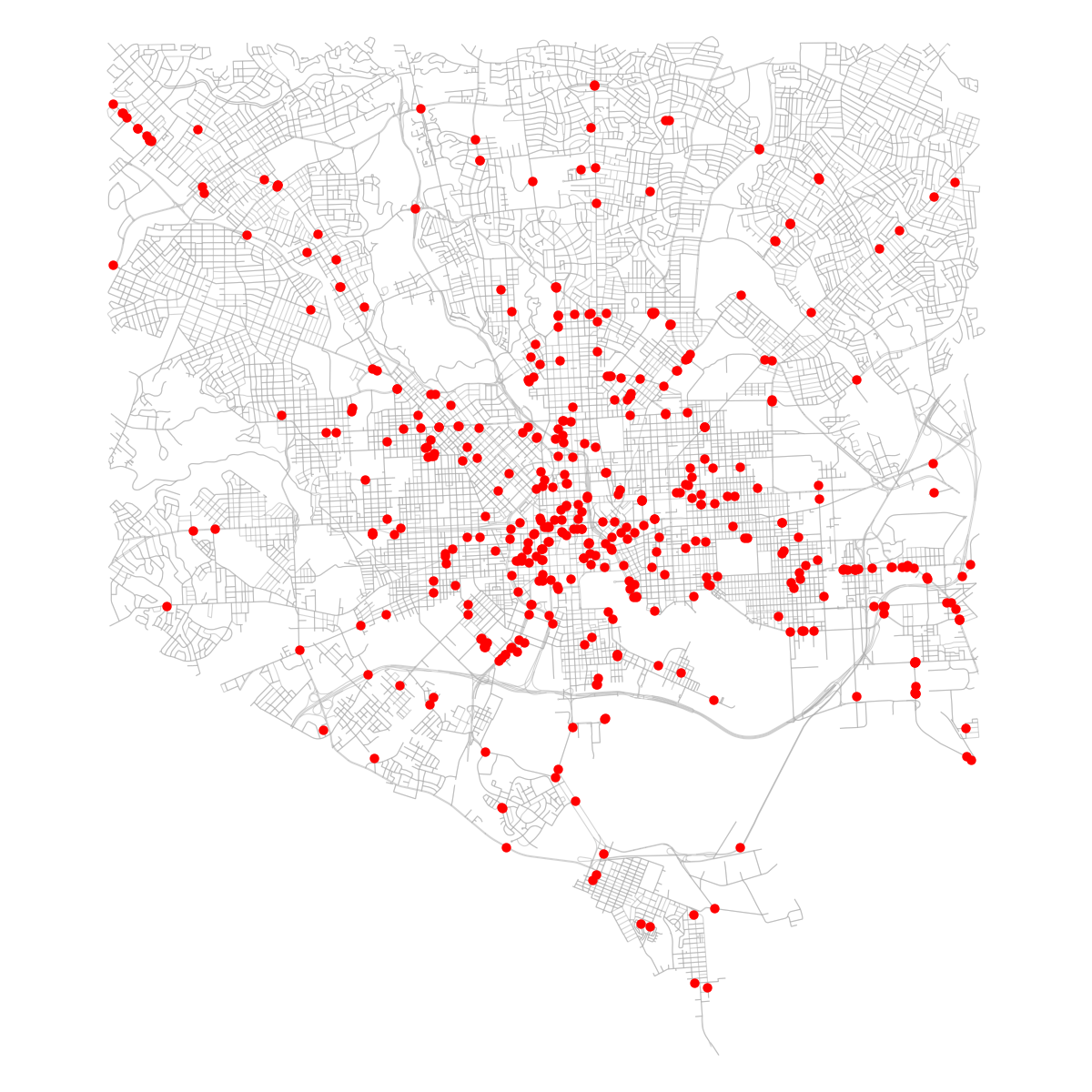} 
    \caption{Baltimore}
  \end{subfigure}
    \begin{subfigure}[t]{.19\linewidth}
    \centering 
    \includegraphics[width=1\textwidth]{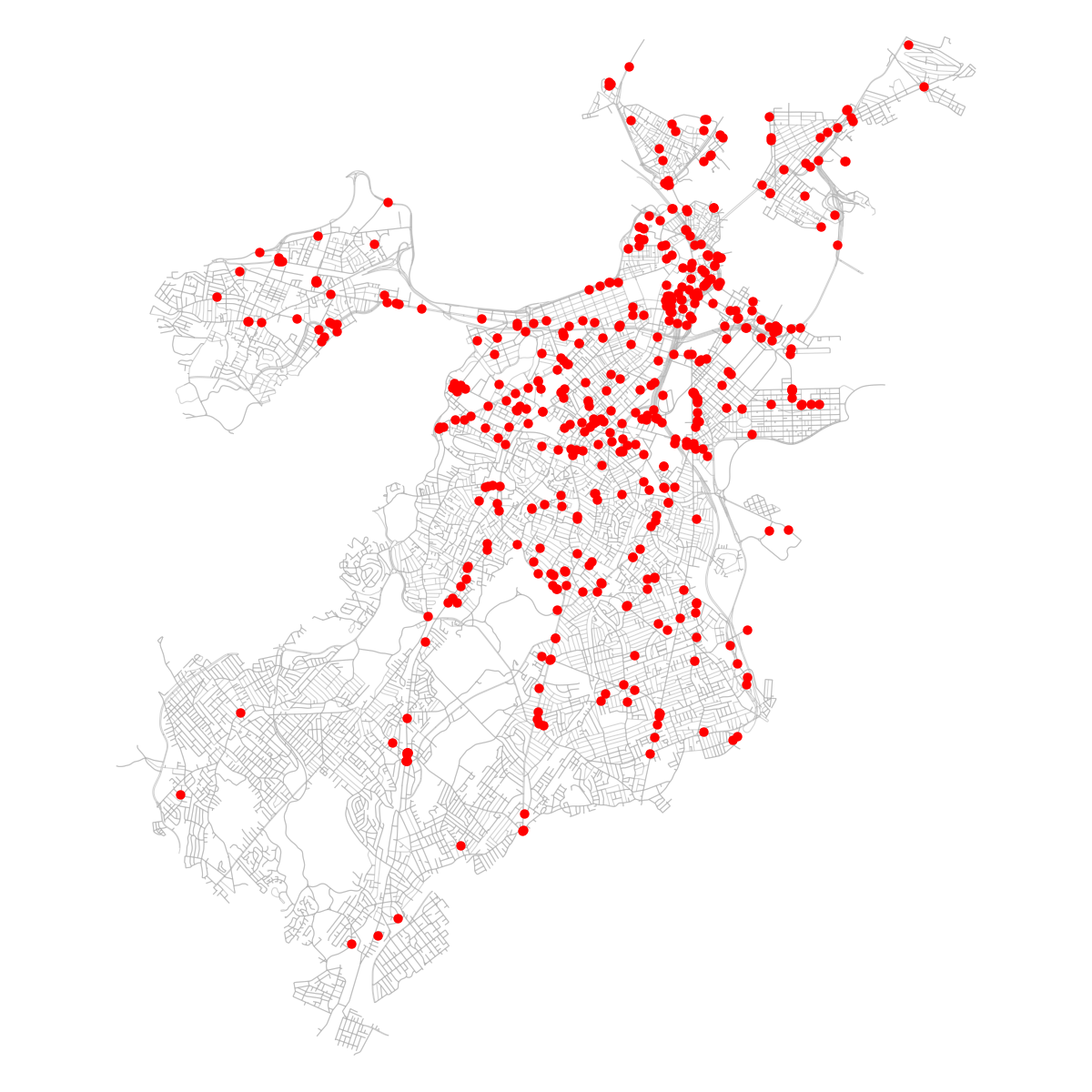} 
    \caption{Boston}
  \end{subfigure}
    \begin{subfigure}[t]{.19\linewidth}
    \centering 
    \includegraphics[width=1\textwidth]{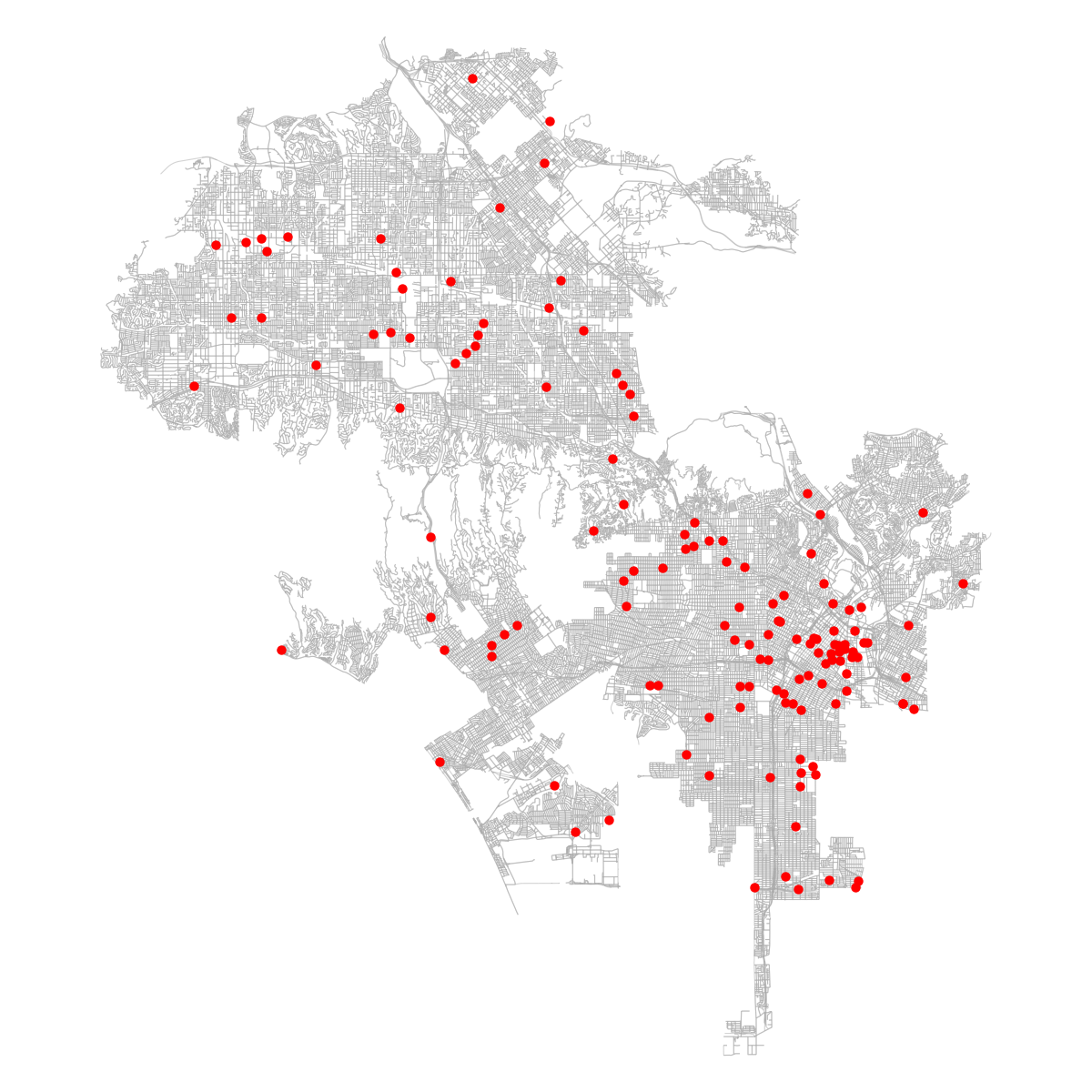} 
    \caption{Los Angeles}
  \end{subfigure}
    \begin{subfigure}[t]{.19\linewidth}
    \centering 
    \includegraphics[width=1\textwidth]{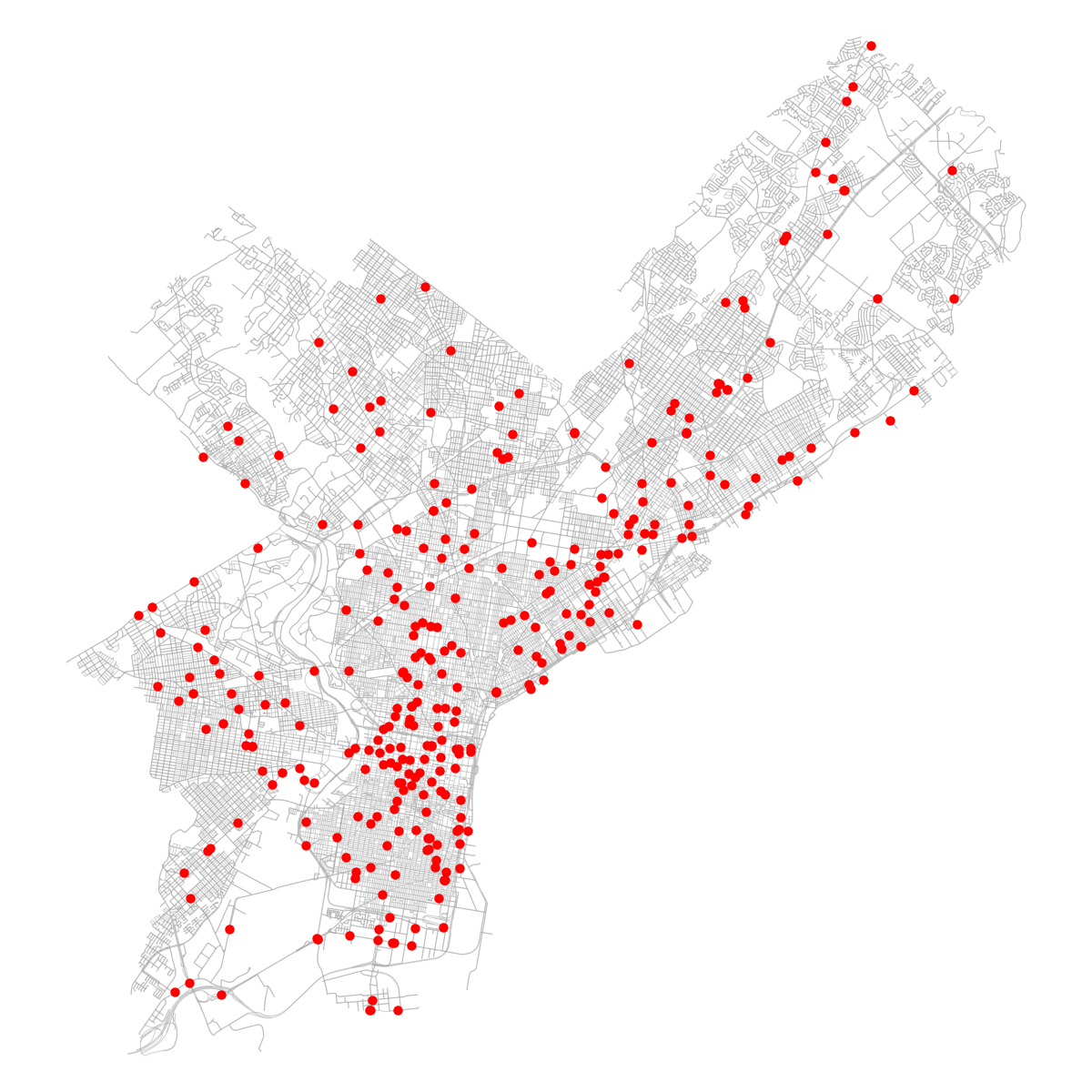} 
    \caption{Philadelphia}
  \end{subfigure}
  
  \caption{Locations of verified cameras in 10 large U.S. cities for the period 2015--2021.
  Densely clustered areas of points indicate regions with high camera density in each city. Camera density varies widely between neighborhoods. Note: Scale varies between cities.
  }
  \label{fig:detections}
\end{figure*}

\subsection{Step 3: Road Network Coverage Estimation}
\begin{figure}[ht!]
  \centering
  \includegraphics[width=0.48\textwidth]{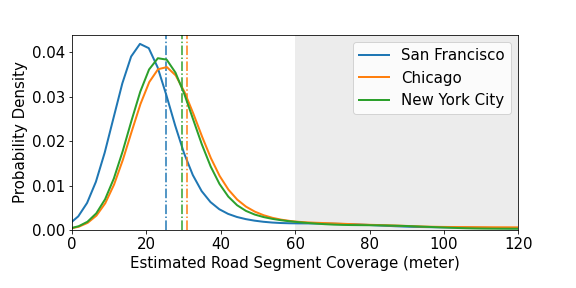} 
  \caption{Distribution of the length of road covered by a  street view image in San Francisco, Chicago and New York City. 
  We limit our analysis to images taken within 30 meters of the nearby buildings---corresponding to 60 meters of street coverage---since we cannot reliably detect cameras at further distances.
  }
  \label{fig:coverage}
\end{figure}

The final step in our procedure is to estimate the fraction of the visible area covered by our randomly sampled images.
To estimate how much of the total length of a city's road network ($D$) has been covered by our sample ($D_{\text{covered}}$), we estimate the average length of road covered by one street view image ($\bar{d}$),
which we can then multiply by the number of images sampled ($N$).

We estimate $\bar{d}$ based on the geometry illustrated in Figure~\ref{fig:geometry}.
Each street view image comes with the exact latitude and longitude of where it was taken. 
Given an image's point location $p$, we find the closest point $p'$ within the nearby buildings' footprint,
and denote the distance between $p'$ and $p$ with $\delta$. 
As discussed above, we chose the street view's heading to be perpendicular to the road orientation,
and restricted it to a 90-degree view.
As a result, we estimate the length of road segment covered by the image to be $d  = 2\delta$.

We repeat this procedure for each sampled street view image.
We remove the relatively small number of images taken at locations more than 30 meters from a building---corresponding to 60 meters of street coverage---since at further distances, cameras become too small to be reliably detected by either humans or computer vision algorithms. 
This results in images that cover about 25--30 meters of street.
For example, the mean road segment covered by an image $\bar{d}$ is 24, 29, and 28 meters in San Francisco, Chicago, and New York City, respectively, as shown in Figure~\ref{fig:coverage}. 

Now, we estimate the proportion of a city's road network covered by our sample as $c = (N \bar{d}) / (2D)$,
where $N$ is the total number of samples for a city (within a given time period) and $D$ is the total length of the city's road network.
Note that the factor of 2 is to account for the fact that
our street view images only cover one of the two sides of a street at any sampled point. 

Finally, putting all the above pieces together, we can estimate the number of cameras $K_{i}$ in city $i$: 
\begin{equation}
 \hat{K}_{i} = \frac{n_{i}}{c_{i}r} \label{eq:estimation},
\end{equation}
where 
$r$ is the recall of our model, 
$n_{i}$ is the number of verified camera detections,
and $c_{i}$ is
the proportion of the road network of a city covered by our sample.

Similarly, we model variance by treating each sampled instance as a draw from a Bernoulli distribution with detection probability $p_{i} = n_{i} / N_{i}$. Assuming that recall and coverage are both exact, we can estimate the standard error of the number of cameras $K_{i}$ as:
\begin{equation}
 \hat{\text{se}}(\hat{K}_{i}) = \frac{\sqrt{N_{i} \cdot p_{i} \cdot (1-p_{i})}}{c_{i}r}.
 \label{eq:variance}
\end{equation}

\section{Results}
Applying the methods described above, we now estimate the total number and spatial distribution of cameras on the road network for all 16 cities. 
In addition, for the U.S. cities, we estimate the prevalence of cameras across city zones, and examine the racial composition of the neighborhoods in which cameras are concentrated.

\subsection{Camera Prevalence}
\begin{table*}[t]
    \centering
    \resizebox{0.9\textwidth}{!}{%
        \begin{tabular}{@{\extracolsep{5pt}} lcc|c|r@{}l|r@{}l} 
            \toprule
            City & Road length (km) & Mean road coverage (m) & \multicolumn{1}{c|}{No. of detections} & Estimated & density  (cameras/km) & Estimated & number of cameras \\
            \midrule 
            Boston & 2,589 & 26 & 516 & 0.63 & (0.03) & 1,600 & (100) \\ 
            New York & 16,362 & 29 & 556 & 0.62 & (0.03) & 10,100 & (400) \\ 
            Baltimore & 3,746 & 30 & 512 & 0.54 & (0.02) & 2,000 & (100) \\ 
            San Francisco & 3,101 & 24 &  398 & 0.52 & (0.03) & 1,600 & (100) \\ 
            Chicago & 10,449 & 30 & 382 & 0.41 & (0.02) & 4,300 & (200) \\ 
            Philadelphia & 6,759 & 29 & 348 & 0.38 & (0.02) & 2,600 & (100) \\ 
            Washington & 3,262 & 33 & 237 & 0.23 & (0.01) & 700 & (50) \\ 
            Milwaukee & 4,899 & 33 & 202 & 0.19 & (0.01) & 900 & (100) \\ 
            Seattle & 5,569 & 29 & 155 & 0.17 & (0.01) & 1,000 & (100) \\ 
            Los Angeles & 21,095 & 29 & 144 & 0.16 & (0.01) & 3,300 & (300) \\
            \midrule
            Seoul & 14,748 & 29 & 869 & 0.95 & (0.03) & 13,900 & (500) \\ 
            Paris & 1,853 & 24 & 590 & 0.76 & (0.03) & 1,400 & (100) \\ 
            Tokyo & 46,688 & 29 & 428 & 0.47 & (0.02) & 21,700 & (1,000) \\ 
            London & 28,907 & 32 & 448 & 0.45 & (0.02) & 13,000 & (600) \\ 
            Bangkok & 34,692 & 29 & 324 & 0.35 & (0.02) & 12,200 & (700) \\ 
            Singapore & 5,794 & 29 & 172 & 0.19 & (0.01)  & 1,100 & (100) \\
            \bottomrule
        \end{tabular} 
    }
    \caption{
    Number of detections and estimates of camera density and total cameras in each city ($N$ = 100,000 images), arranged in descending order of camera density. Standard errors are listed in parentheses. 
    We impute a mean road coverage of 29m for cities without adequate building footprint data.
    }
    \label{tab:results}
\end{table*}
\begin{figure}[ht!]
  \centering
  \definecolor{pre}{HTML}{00BFC4}
  \definecolor{post}{HTML}{F8766D}
  
  \includegraphics[width=0.45\textwidth]{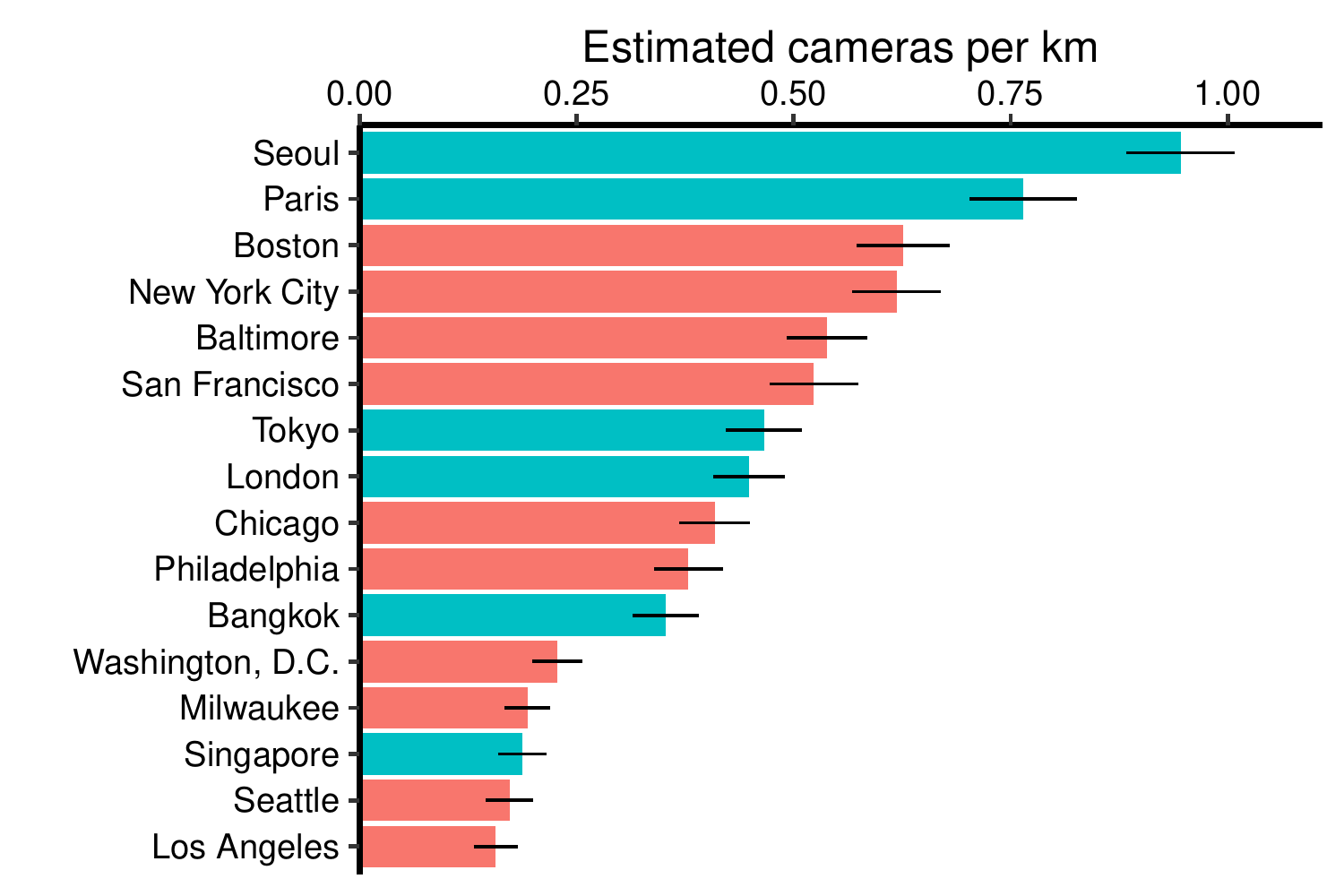}
  \caption{%
  Estimated camera density (cameras per km) for {\color{post}$\blacksquare$} 10 large U.S. cities and {\color{pre}$\blacksquare$} 6 other major cities.
  Horizontal lines indicate 95\% CIs for the estimate. 
  }
  \label{fig:result-density}
\end{figure}
Table \ref{tab:results} shows the number of identified cameras for each city---after human verification---along with point estimates and 95\% confidence intervals for camera density and for the total number of cameras, following Eqs.~\eqref{eq:estimation} and \eqref{eq:variance}.
The same density estimates are also depicted in descending order in Figure \ref{fig:result-density}. 
We find that camera density varies widely between cities: For example, Boston and New York City, the U.S. cities with the highest camera density, have almost four times as many cameras per kilometer than Seattle and Los Angeles.\footnote{%
Our computer vision model was trained on San Francisco data, and so it is possible that the camera identification rate in San Francisco is inflated due to over-fitting. 
We note, however, that while model precision varies between cities, Philadelphia and Boston both have higher precision than San Francisco, which suggests our model does indeed transfer well across contexts. 
}
We note that our estimates exclude indoor cameras, 
as well as outdoor cameras not captured by street view images.
Perhaps due to these limitations, our estimate of 10,100 cameras in New York City is lower than the 18,000 cameras that the NYPD reportedly has access to~\cite{nypd_camera_count}.

\subsection{Camera Placement}
\begin{figure}[ht!]
  \centering
  \includegraphics[width=0.4\textwidth]{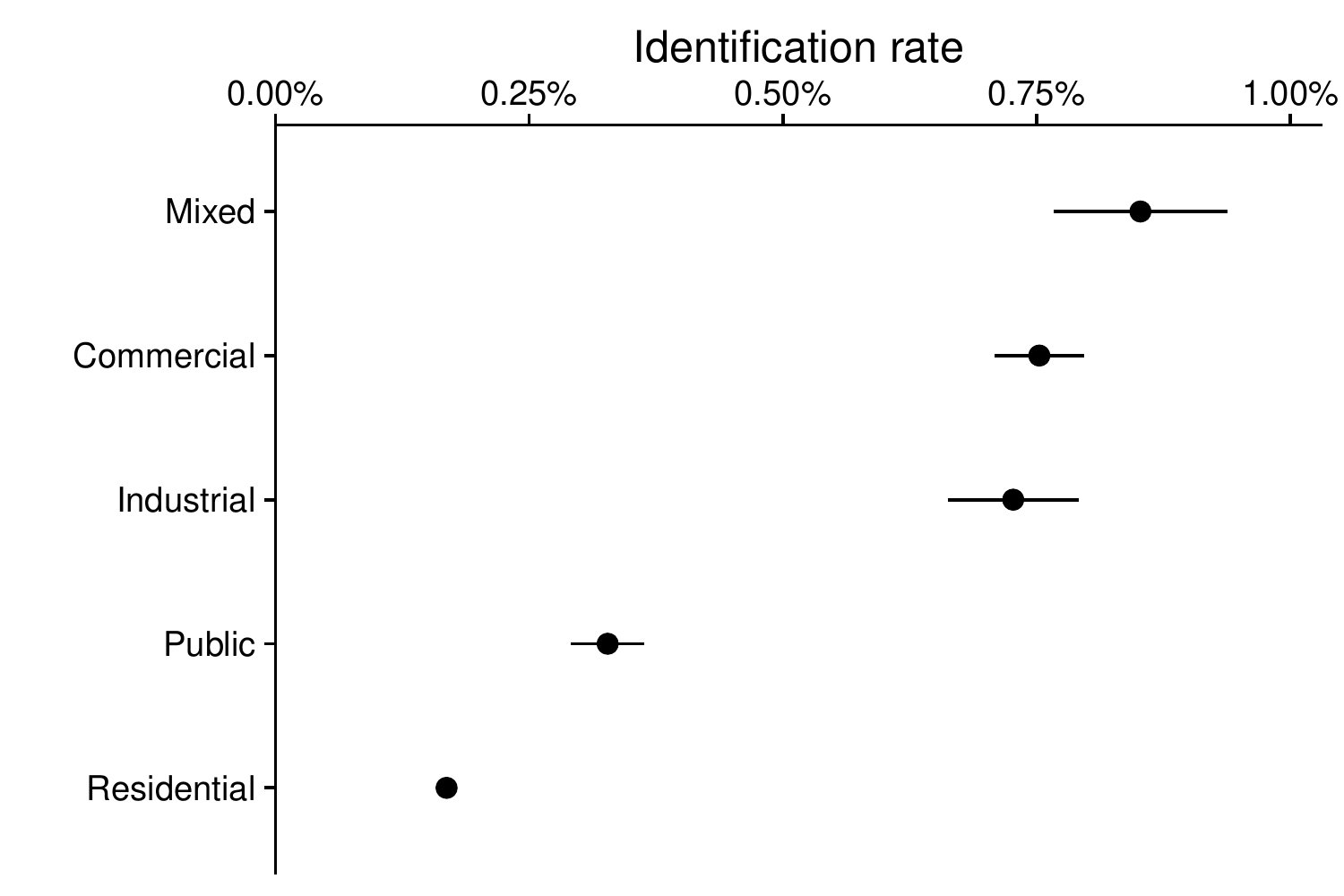}
  \caption{
  Camera identification rates (cameras per image) for different zoning designations across 10 large U.S. cities for the period 2015--2021. 
  Horizontal lines indicate 95\% CIs for the estimate. 
  Cameras are more likely to be detected in mixed, industrial, and commercial zones than in public and residential zones. 
  }
  \label{fig:result-zone}
\end{figure}
\begin{figure}[!ht]
  \centering
  \includegraphics[width=0.45\textwidth]{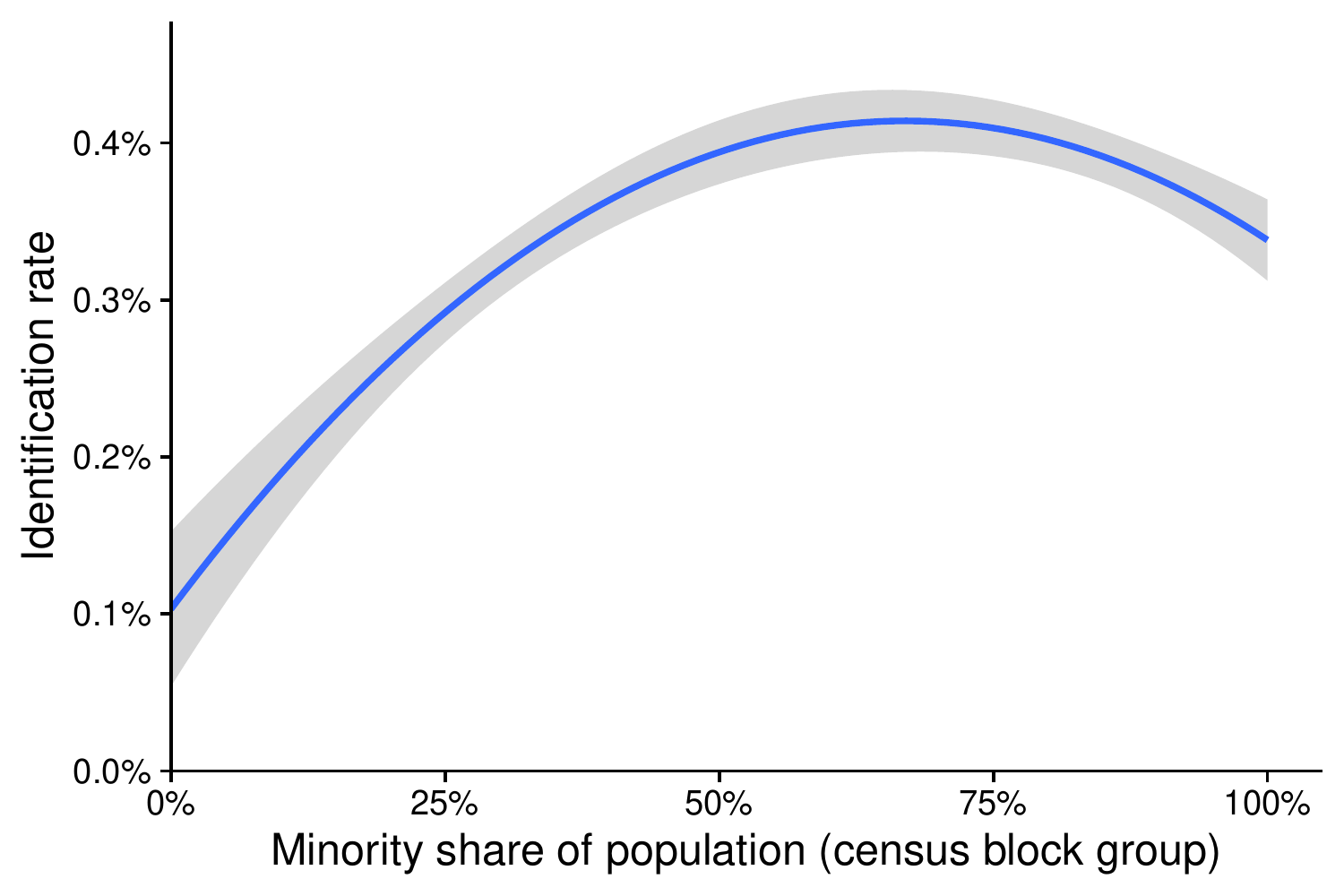} 
  \caption{
  The relationship between camera identification rate (cameras per image) and minority share of population, aggregated across 10 large U.S. cities for the period 2015--2021. 
  The blue line shows a regression fit to the data, and indicates that
  cameras are more likely to be detected in majority-minority neighborhoods than in predominantly non-Hispanic white neighborhoods. 
  }
  \label{fig:result-race}
\end{figure}

The detection maps in Figure~\ref{fig:detections} show that cameras are not distributed uniformly across a city. Despite sampling uniformly over the road network, we find densely covered regions in each city, representing neighborhoods with a high concentration of cameras. 
We examine these patterns in more detail for the 10 U.S. cities we consider, analyzing the rate of (verified) camera identifications per street image across zoning designations and neighborhood racial composition.

Figure~\ref{fig:result-zone} shows the camera identification rate for different zoning designations aggregated over the 10 U.S. cities we analyze.
We find that images from mixed, industrial, and commercial zones are more likely to contain an identified camera than images from public (such as parks and other public facilities) and residential areas. 
For example, the identification rate in mixed zones (2.1\%) is more than three times the rate in residential zones (0.6\%).
This pattern holds for the majority of our chosen cities. 

To compute the camera identification rate, we assigned each sampled point to the zoning designation of the closest parcel of land. To do so, we collected land use and zoning designation data for all 10 cities, and then standardized the zoning code into one of the following five categories: mixed, industrial, commercial, public, and residential. Zones with codes that represent planned development or did not clearly fit into the aforementioned categories are labeled as unknown and
omitted in the following analysis. We find that 60\% of sampled points are classified as residential, and unknown codes comprise less than 3\% of sampled points. 

We next examine the relationship between camera identification rate and the share of residents in the surrounding area that identify as belonging to a minority racial or ethnic group, 
aggregated over our 10 U.S. cities. 
To compute this relationship, we assigned each sampled image to the minority proportion of the census block group in which it is located, as estimated by the 2018 American Community Survey.
For purposes of this analysis, we define ``minority'' as comprising those individuals who identify either as Hispanic (regardless of their race) or who do not identify as white.

Figure \ref{fig:result-race} shows the results.
The blue line is a regression (with both linear and quadratic terms) fit to the data, and indicates that an increase in the share of minority residents in a neighborhood is associated with an increase in camera identification rate. 
For example, the identification rate in census blocks with a 50\% minority share (0.38\%) is twice as high as in those blocks with a 10\% minority share (0.2\%).
We see qualitatively similar results with higher-order polynomial fits.

\begin{table}[t] 
    \newcommand{\dtoprule}{\specialrule{1pt}{0pt}{0.4pt}%
                \specialrule{0.3pt}{0pt}{\belowrulesep}%
                }
    \newcommand{\dbottomrule}{\specialrule{0.3pt}{0pt}{0.4pt}%
                \specialrule{1pt}{0pt}{\belowrulesep}%
                }

    \centering 
    \resizebox{0.4\textwidth}{!}{
    \begin{tabular}{lc} 
    \dtoprule
    & \multicolumn{1}{c}{Identification Rate} \\ 
    \midrule
      Public & 0.0014${}^{***}$ (0.0002) \\ 
Commercial & 0.0055${}^{***}$ (0.0002) \\ 
Industrial & 0.0053${}^{***}$ (0.0002) \\ 
Mixed & 0.0060${}^{***}$ (0.0003) \\ 
Percentage minority & 0.0059${}^{***}$ (0.0009) \\ 
Percentage minority${}^2$ & -0.0044${}^{***}$ (0.0008) \\
    \midrule
    Observations & \multicolumn{1}{c}{787,418} \\ 
    \dbottomrule
    \textit{Note:}  & \multicolumn{1}{r}{$^{*}$p$<$0.1; $^{**}$p$<$0.05; $^{***}$p$<$0.01} \\ 
    \end{tabular}
    }
    
    \caption{
      Coefficients and standard errors for a linear probability model of camera identification as a function of city, zone, and minority share (with a quadratic term), fit
      on sampled points in the period 2015--2021. 
      The reference group for zone is residential neighborhoods. 
      }
    \label{tab:regression} 
\end{table} 
The observed concentration of cameras in majority-minority neighborhoods persists even after adjusting for zone category.
Specifically, Table~\ref{tab:regression}
shows the results of a linear probability model
that predicts camera detections as a function of city, zone, and racial composition---where we again use a quadratic term to account for the curvature seen in Figure~\ref{fig:result-race}. 
The fitted model confirms that camera identifications increase with the minority share of residents, plateauing at approximately 60\% share and then remaining relatively flat, as in Figure~\ref{fig:result-race}.
It is unclear what is driving the apparent concentration of cameras in high minority neighborhoods.
However, regardless of the underlying mechanism, these results point to the potential impacts that video surveillance can have on communities of color.

\section{Discussion and Conclusion}
By applying computer vision, human verification, and statistical analysis to large-scale, geo-tagged image data, 
we have---for the first time---estimated the number and spatial distribution of outdoor surveillance cameras 
in 16 major cities around the world. 
Further, the approach we have developed has the potential to scale to even more cities across the country and the world, providing a new perspective on the state of video surveillance.

In the 16 cities we analyzed,
we found considerable variation in the estimated number of surveillance cameras.
Among U.S. cities, our analysis also shows that cameras are more likely to be found in industrial, commercial, and mixed zones as compared to residential areas. 
Finally, even after adjusting for zone category, 
we find a greater concentration of cameras in 
majority-minority neighborhoods, highlighting the need to carefully consider the potential disparate impacts of surveillance technology on communities of color.
    
While our computational approach is able to provide a novel quantitative perspective into the state of surveillance, it 
is still subject to several important limitations, which we outline below.
First, our method relies on being able to see cameras from the street, and, more specifically, from street view images. 
Indoor cameras, as well as outdoor cameras obscured from view,
are not counted by our estimation pipeline. 
Further, due to the limited resolution of street view images, small cameras---such as increasingly popular doorbell cameras---are difficult to detect by either humans or algorithms.
Higher resolution and higher coverage image data could mitigate these issues in the future.
However, our current results likely underestimate the density of cameras in a city.

Second, our human annotators may not perfectly label cameras in the candidate images selected by the model, skewing our final estimates.
For example, it is possible that they rule out an actual camera (leading to an undercount) or, conversely, that they report a camera that is not in fact there (leading to an overcount). 
To minimize these errors, every candidate image is independently labeled by three human annotators, 
but at least some errors likely remain.

Third, errors in the estimated recall of our computer vision model---and, similarly, errors in the estimated coverage of our images---can bias our final estimates.
Estimating city-specific model recall is particularly challenging, as it requires city-specific labeled datasets.
In our analysis, we thus estimated recall for a single city, San Francisco, in which the locations of some surveillance cameras had already been compiled, which we then apply to other jurisdictions.
Further, our variance estimates treat the recall and coverage as known quantities. 
Accounting for errors in their measurement would increase the variance of our final estimates.

Finally, our method does not provide any information about the cameras other than what can be inferred from their appearance.
For example, we cannot determine whether identified cameras are decoys,
are malfunctioning, or otherwise are not in use.
We likewise cannot always tell who owns the cameras (e.g., a government agency or a private citizen), who has access to the video,
and whether the camera footage is stored.
All of these factors are critical in assessing the downstream consequences of video surveillance.
Although difficult, future work may be able to answer some of these questions by conducting 
a more intensive audit of a sample of the identified cameras.

Despite these limitations,
we believe our approach and results constitute an important step toward understanding the use of surveillance technology across the world.
More broadly, our 
general statistical estimation pipeline can be extended and applied to characterize the prevalence and spatial distribution of a variety of other city elements detectable from street images.
Looking forward, we hope this work spurs further theoretical and empirical research at the intersection of computer vision, urban computing, and public policy. 

\section*{Publication Note}
This version of the paper is updated from our original publication in two important respects.
First, we now credit
\citet{turtiainen2020cctv} both for creating a state-of-the-art camera detection model and for suggesting that computer vision could, in theory, be applied to street view data to map surveillance cameras.
We were aware of their work when initially conducting our research, but we unfortunately failed to include a citation to their paper.
We thank \citeauthor{turtiainen2020cctv} for bringing this to our attention and we apologize for the omission.
Second, we discovered a coding error in our image sampling strategy that corrupted our analysis of camera density over time. We have now removed the results of that analysis.

\balance
\bibliographystyle{ACM-Reference-Format}
\bibliography{ref}

\end{document}